\documentclass[runningheads]{llncs}
\usepackage{graphicx}
\usepackage{amssymb,amsmath,amsfonts,amstext, marvosym, graphics, psfrag, bm}
\usepackage[ruled,linesnumbered]{algorithm2e}
\usepackage{subfigure}
\usepackage{float} 
\usepackage{color}
\usepackage{comment}
\newcommand\mynote[1]{\textcolor{black}{#1}}

 \usepackage{caption}
\graphicspath{{newPhancomparison/}{invivo_FLAIRcomparisonfull/}{compare2RAKIfigures/}}

\begin{document}
\title{APIR-Net: Autocalibrated Parallel Imaging Reconstruction using a Neural Network}

%
\titlerunning{APIR-Net}

\author{Chaoping Zhang\inst{1} \and
Florian Dubost\inst{1} \and
Marleen de Bruijne\inst{1,2} \and
Stefan Klein\inst{1} \and
Dirk H.J. Poot\inst{1}}
\authorrunning{C. Zhang et al.}
%
\institute{Biomedical Imaging Group Rotterdam, Erasmus MC, Rotterdam, The Netherlands \and
Department of Computer Science, University of Copenhagen,
Copenhagen, Denmark}

\maketitle              
\begin{abstract}
Deep learning has been successfully demonstrated in MRI reconstruction of accelerated acquisitions. However, its dependence on representative training data limits the application across different contrasts, anatomies, or image sizes. To address this limitation, we propose an unsupervised, auto-calibrated k-space completion method, based on a uniquely designed neural network that reconstructs the full k-space from an undersampled k-space, exploiting the redundancy among the multiple channels in the receive coil in a parallel imaging acquisition. To achieve this, contrary to common convolutional network approaches, the proposed network has a decreasing number of feature maps of constant size. In contrast to conventional parallel imaging methods such as GRAPPA that estimate the prediction kernel from the fully sampled autocalibration signals in a linear way, our method is able to learn nonlinear relations between sampled and unsampled positions in k-space. The proposed method was compared to the start-of-the-art ESPIRiT and RAKI methods in terms of noise amplification and visual image quality in both phantom and in-vivo experiments. The experiments indicate that APIR-Net provides a promising alternative to the conventional parallel imaging methods, and results in improved image quality especially for low SNR acquisitions.
\end{abstract}
\section{Introduction}
Magnetic resonance imaging (MRI) provides versatile contrast information for clinical diagnosis. However, its long scan time remains a limitation. To reduce scan time, parallel imaging \cite{Pruessmann1999,griswold2002generalized} has been proposed to reconstruct subsampled k-spaces acquired by multi-channel coils and is widely used in clinic. Recently, deep learning was also demonstrated to enable fast imaging with the reconstruction model trained on representative data \cite{hammernik2018learning,wang2016accelerating,zhu2018image}. 

Despite the current success of deep learning in MRI reconstruction, most methods are size, contrast, or anatomy specific. Also they depend on the corresponding training data, and may create inaccurate reconstruction for features not seen in training data. Recurrent inference machines have been introduced to iteratively reconstruct heterogeneous raw MRI data with different anatomies and acquisition settings \cite{lonning2019recurrent}. However, training data is still needed and influences reconstruction performance. A database-free deep learning approach for fast imaging (RAKI) was proposed for parallel imaging reconstruction \cite{akccakaya2019scan}. It learns the prediction kernel with an artificial neural network from fully sampled autocalibration signals (ACS) and subsequently uses the learned kernel to predict the unsampled signals. In this method, the nonlinear estimation of the prediction kernel enables improved noise resilience compared to the linear GRAPPA method.

In this work, we propose a different unsupervised k-space completion method for parallel imaging, called Autocalibrated Parallel Imaging Reconstruction using a Neural Network (APIR-Net). Contrary to RAKI which as a 2D method predicts the unsampled signals for a 2D k-space using prediction kernels learned from the ACS signals, APIR-Net predicts all signals of a 3D full k-space from the subsampled k-space utilizing all sampled signals, including ACS signals and beyond. 

Most image based neural network architectures use downsampling (and subsequent upsampling) operators with increasing number of feature maps to force the network to use higher level image features. This assumes that such high level features are present at rather small scales in the images. In k-space small scale features represent large scale image features and hence such high level features are less likely to be present, yet preservation of small scale information is essential. On the other hand, for MRI using a multi-channel receive coil, signal redundancy exists among the channels. Hence, inspired by, but in contrast to, the U-net architecture, APIR-Net decreases the number of feature maps while preserving their size throughout all layers.

To improve the computational efficiency, we propose to train APIR-Net in a hierarchical process, starting from a small portion of k-space in the center region until the full size k-space. The network trained at a lower level provides initialized weights for a subsequent higher level's training. The performance of APIR-Net was evaluated with phantom and in-vivo acquisitions with comparison to GRAPPA \mynote{\cite{griswold2002generalized}}, ESPIRiT \cite{uecker2014espirit}, and RAKI \mynote{\cite{akccakaya2019scan}} methods.

\begin{figure}[H]
\centering
 \subfigure[$M_{sampled}$]{\includegraphics[width = 0.2 \textwidth]{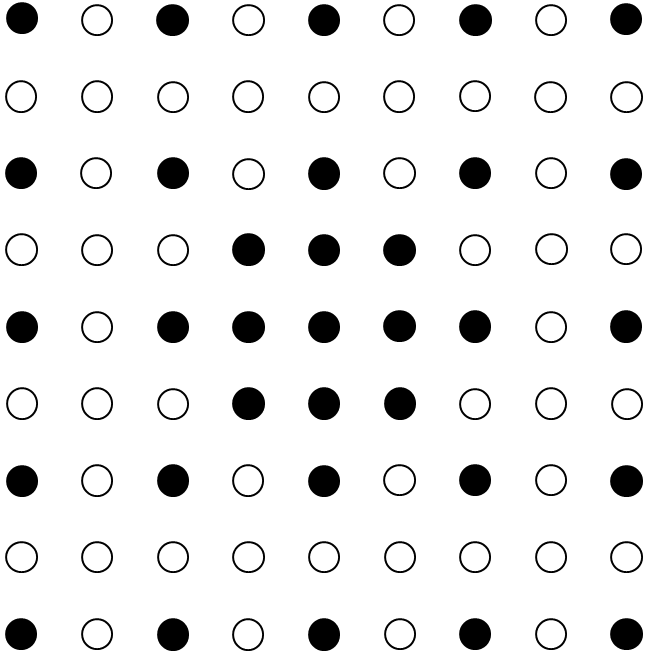}}~~~~~
  \subfigure[$M_{pattern}$]{\includegraphics[width = 0.2 \textwidth]{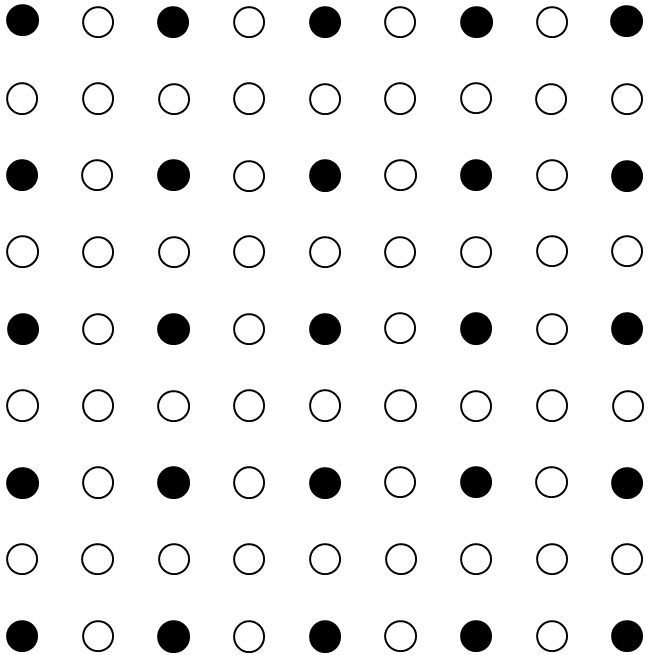}}~~~~~
  \subfigure[$M_{acs}$]{\includegraphics[width = 0.2 \textwidth]{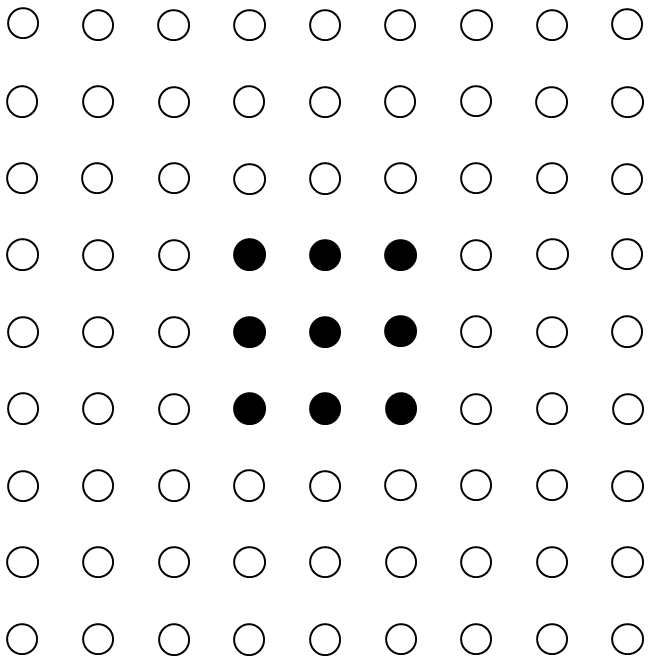}}~~~~
  \includegraphics[width = 0.14 \textwidth]{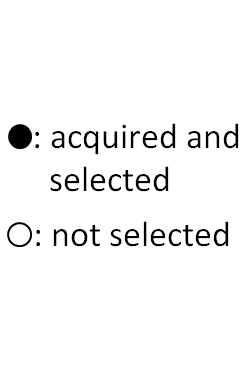}
\caption{An example of the masks of all sampled positions $M_{sampled}$, regularly subsampled positions $M_{pattern}$, and the ACS positions $M_{acs}$ in PE directions.}
\label{figure_mask_ksp}
\end{figure}

\section{Methods}
\subsection{Conventional Parallel Imaging Reconstruction \cite{griswold2002generalized}}
In a parallel imaging acquisition with a multi-channel receive coil, k-space is subsampled regularly with a small fully sampled ACS region in k-space center. We represent the masks of all sampled positions, regularly subsampled positions, and the ACS positions in phase encoding (PE) directions as $M_{sampled}$, $M_{pattern}$, and $M_{acs}$, respectively, as illustrated in Fig.~\ref{figure_mask_ksp}. In GRAPPA, the unsampled signals are predicted as 
\begin{equation}
\begin{aligned}
{S_{predicted}} = n\circledast ( S \circ M_{pattern} ),
\end{aligned}
\label{GRAPPA_equation}
\end{equation}
where $\circledast$ represents convolution operation and $\circ$ represents pixelwise multiplication. $S$ represents all signals in k-space. $n$ is the prediction kernel that is trained on the ACS region $M_{acs}$ by the least squares fitting as
\begin{equation}
\begin{aligned}
 \hat{n}=\operatorname*{arg\min}_{n}\|(S - n\circledast ( S \circ M_{pattern} )) \circ M_{acs}  \|_2^2 + \lambda \|I n\|_2^2,
\end{aligned}
\label{GRAPPAKernel_equation}
\end{equation}
where $\lambda$ is a scalar and $I$ is the identity matrix. The second term in equation \ref{GRAPPAKernel_equation} represents Tikhonov regularization, and actual implementations may add terms with shifted versions of $M_{pattern}$. 

The final interpolated k-space is computed as
\begin{equation}
S_{final} = S \circ M_{sampled} + S_{predicted} \circ (1-M_{sampled}),
\label{ksp_interpolated}
\end{equation}
and the final image $X$ is computed as
\begin{equation}
X = \sqrt {\frac{1}{C}{\sum\limits_c |{\textrm{iFFT}_c( S_{final} )}|^2}},
\label{ImgRecon_RMS}
\end{equation}
where $C$ is the number of channels and $\textrm{iFFT}_c$ is the 3D inverse Fourier transform on channel $c$.

\subsection{APIR-Net Reconstruction}
\subsubsection{Problem Formulation} Instead of explicitly computing the convolution kernel $n$ and applying it over the full k-space as in GRAPPA \cite{griswold2002generalized} or RAKI \cite{akccakaya2019scan}, our approach recovers the full k-space by training of a deep convolutional neural network:

\begin{equation}
\hat{\theta} = \operatorname*{arg\min}_{\theta} \left \| \left ( A_{\theta}(S \circ M_{pattern}) - S\right ) \circ M_{sampled}\right \|_2^2,
\label{APIR_Net_equation}
\end{equation}

\noindent where $A_{\theta}$ is the function to predict the full k-space from $S \circ M_{pattern}$ and is parametrized by $\theta$ as a neural network. With the input equal to $S \circ M_{pattern}$, only the regularly subsampled signals are effectively fed into the network, whereas the loss is computed on all sampled positions $M_{sampled}$. The final image $X$ is reconstructed by equation \ref{ImgRecon_RMS} with $S_{final} =A_{\hat{\theta}}(S \circ M_{pattern})$.

\vspace{-0.1cm}
\subsubsection{Network Architecture} 
\begin{figure}[t]
\includegraphics[width=\textwidth]{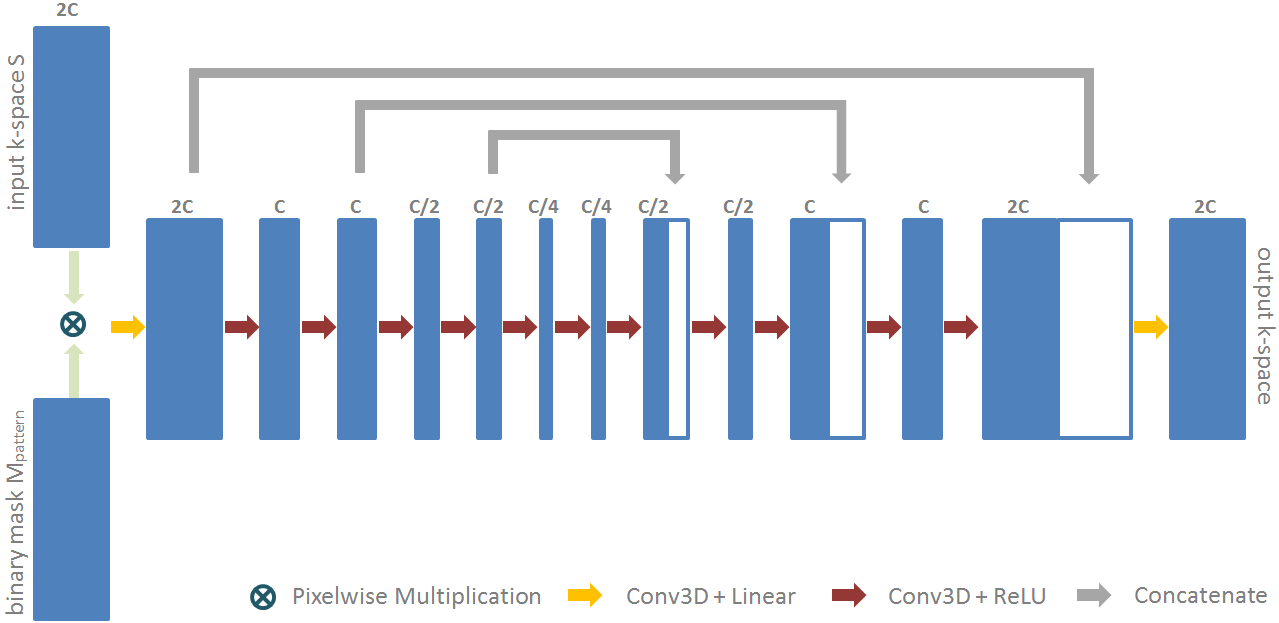}
\caption{The neural network architecture of APIR-Net.} 
\label{fig_Net}
\end{figure}

The detailed network architecture is shown in Fig.~\ref{fig_Net}. The real and imaginary components of the complex k-space are concatenated in channel dimension, resulting in an input with $2C$ channels. Similarly, the output with $2C$ channels is finally converted to a $C$-channel complex-valued k-space. 
As the multi-channel k-space in a parallel imaging acquisition is highly correlated across the channel dimension, in APIR-Net, the layers have progressively reduced number of channels, or feature maps, as the depth of the encoder increases, while the size of each feature map remains unchanged.
The first and last convolutional layers use a kernel size $5\times5$ followed by the linear activation function. The remaining convolutional layers are with kernel size of $3\times3$ followed by the ReLU activation function. Periodic padding is used for all convolutional layers, such that the border of the input for each convolutional layer is padded from the opposite border to maintain the size of output equal to the input. 

\subsubsection{Training} 
The network is trained in a hierarchical way. This is motivated by the main aspect that allows parallel imaging reconstruction: the differences in coil sensitivity of the multi-channel coil. In image domain the coil sensitivity is multiplicative in k-space this appears as translation invariant convolution. Hence, to accelerate training we start with the ACS part and progressively increase the size until the full k-space is included. 
Details of the training are provided in Algorithm \ref{algo_hier_train}. 

\begin{algorithm}[hbt!]
 \SetKwInOut{Input}{Input}\SetKwInOut{Output}{Output}
 
 \Input{$S$, $M_{sampled}$, $M_{pattern}$, $L\leftarrow$ number of levels in the hierarchical training,
$\bm{k_l}\in\mathbb{Z}^3\leftarrow$ position vector indicating the k-space region being used in training of level $l$}
 \Output{$\hat{\theta}$}
 Normalize magnitude $|S \circ M_{sampled}|$ to [0,1]\\
 Initialize $\hat{\theta}_0$ randomly with a uniform distribution within [-0.05, 0.05]\\ 
 \For{$l\leftarrow 1$ \KwTo $L$}
 {
  $S_{in,l} = S(\bm k_l) \circ M_{sampled}(\bm k_l)$, $M_{in,l}=M_{pattern}(\bm k_l)$, $M_{out,l}=M_{sampled}(\bm k_l)$\\
  $\hat{\theta}_l = \operatorname*{arg\min}_{\hat{\theta}} \left \| \left ( A_{\theta}(S_{in,l}\circ M_{in,l}) - S_{in,l}\right )  \circ M_{out,l} \right\|_2^2$ starting from $\theta=\hat{\theta}_{l-1}$ until convergence
 }
 \KwRet{$\hat{\theta} = \hat{\theta}_L$}
 \caption{The proposed hierarchical training method.}
 \label{algo_hier_train}
\end{algorithm}%

\section{Experiments}
\subsection{Evaluation with Phantom Acquisition}
The 3D k-space of the ACR-NEMA MRI Phantom was fully acquired with fast spin echo sequence using a 3T GE Discovery MR750 scanner and an eight-channel birdcage-like receive brain coil (8HRBRAIN). The scan parameters include \mynote{repetition time $(TR)=2800ms$, echo train length $(ETL)=60$, bandwidth $(BW)=83.33kHz$, field of view $(FOV)=20.5\times20.5\times20.5 cm^3$,} $\text{Matrix~size} = 192\times192\times192$. 
The cylindrical phantom was placed axially in the coil array, with S/I as frequency encoding (FE) direction. 

Reconstructions with GRAPPA \mynote{\cite{griswold2002generalized}}, ESPIRiT \cite{uecker2014espirit}, $l_1$-ESPIRiT (ESPIRiT integrating the regularization of the $l_1$-norm with a sparsity transform) \cite{uecker2014espirit} and APIR-Net were performed on a retrospectively subsampled k-space from the full acquisition with a subsampling factor of $2\times2$ in two PE directions. The k-space center with the size [$25\times25$] in [PE1, PE2] was fully sampled as the ACS region. To highlight the noise amplification suppression capability of both methods, simulated Gaussian noise was added to the acquired positions in the subsampled k-space. 

For GRAPPA, a convolution kernel size of [5, 9, 9] in [FE, PE1, PE2] directions was selected from a range of options to obtain an optimal reconstruction. 
To fairly compare GRAPPA we reconstructed both without Tikhonov regularization ($\lambda = 0$) as well as with a value of $\lambda$ for which aliasing artifacts started to appear. 

ESPIRiT reconstruction was performed using implementation from the BART toolbox \cite{uecker2015berkeley}. In the reconstruction, the eigenvector maps for the first two eigenvalues were used. For $l_1$-ESPIRiT reconstruction, a regularization term of $l_1$-norm with wavelet transform was used. The strength of the regularization was selected towards a low noise level while avoiding visually obvious blurriness or artifacts ($r=0.01$).

In APIR-Net reconstruction, the training was converged with sufficient number of epochs. The settings of the hierarchical training are heuristically determined and are shown in Table~\ref{tab_levels}. Adam was used as optimizer ($\beta_1 = 0.9$, $\beta_2 = 0.99$, $\epsilon = 10^{-20}$) and no regularization was used in APIR-Net. 

The computation time was around 5 minutes (min) for GRAPPA and around 10 min for ESPIRiT with the CPU Intel Xeon E5503, and was around 120 min for APIR-Net with the GPU NVIDIA GeForce GTX 1080Ti and the CPU Intel Core i7-8700.

\begin{table}[hbt!]
\caption{The settings for each level of the hierarchical training in APIR-Net.}
\centering
\begin{tabular}{c|c|c|c|c}
\hline
        & Size (phantom) & Size (in-vivo) & \begin{tabular}[c]{@{}l@{}}Initial learning rate\end{tabular} & \begin{tabular}[c]{@{}l@{}}Number of epochs\end{tabular} \\ \hline
Level 1 & 32x32x32       & 16x32x32       & 0.001                                                            & 10000                                                                                                                               \\
Level 2 & 48x48x48       & 112x84x64      & 0.0001                                                           & 5000                                                                                                                                \\
Level 3 & 96x96x96       & 224x164x126    & 0.00005                                                          & 1000                                                                                                                                \\
Level 4 & 192x192x192    & 224x224x178    & 0.00005                                                          & 500                                                                                                                                 \\ \hline
\end{tabular} 
\label{tab_levels}
\end{table}

The mean square error (MSE) of the phantom region in the reconstructed image with regard to the reference image was computed for each method. The reference image for MSE computation was reconstructed by the root mean squares of the inverse Fourier transform on the fully acquired k-space.

The noise amplification factor was computed with the pseudo multiple replica method \cite{robson2008comprehensive} with 50 iterations by adding Gaussian white noise to the acquired positions in the subsampled k-space. The magnitude level of the simulated noise was the same for all replica. 

\subsection{Comparison to RAKI}
As the current version of RAKI is a 2D method, we performed a separate experiment for comparison. The implementation of RAKI was kindly provided by the authors of \cite{akccakaya2019scan}. To fit a 2D reconstruction method, the same 3D k-space was first fourier transformed in the FE direction to obtain k-spaces of 2D axial (PE1$\times$PE2) slices, which contains the most variation of the coil sensitivity. A single slice was extracted and was further subsampled by a factor of 3 in the first PE direction, with an ACS region of 25 lines. This data was reconstructed by RAKI, APIR-Net, ESPIRiT, $l_1$-ESPIRiT ($r=0.01$), and GRAPPA, where (obviously) 2D Convolution kernels of otherwise identical size were used in APIR-Net (2D APIR-Net). To investigate the influence of increasing training size we additionally reconstruct with APIR-Net the 3D k-space identically subsampled by a factor 3 in the first PE direction.

\subsection{Evaluation with In-vivo Acquisitions}
To evaluate the proposed method with in-vivo acquisitions, a brain scan from one volunteer with FLAIR contrast was performed with the same scanner and coil as the phantom acquisition. This study was approved by our Institutional Review Board and informed consent was obtained from the volunteer. The prospectively subsampled k-spaces skipped the corners in the PE plane. The subsampling factors are [2, 3] in [PE1, PE2] directions. The ACS region with size [$25\times25$] in [PE1, PE2] was additional fully acquired as well. Other settings include $TR=5000ms$, \mynote{inversion time} $(TI)=1700ms$, $ETL=60$, $FOV=22.4\times22.4\times17.8cm^3$, $\text{Matrix~size} = 224\times224\times178$, $BW=41.67kHz$. The effective scan time was 3.95 min. 

Similar to the experiments with phantom data, Gaussian noise was added to the acquired positions in k-spaces to investigate the noise amplification suppression capability of the methods. The settings of the hierarchical training for APIR-Net are shown in Table~\ref{tab_levels} as well. 
For APIR-Net reconstruction, prediction using network weights trained on different levels of the hierarchical training was also performed. Besides APIR-Net, reconstructions with GRAPPA, ESPIRiT (eigenvectors of the first two eigenvalues used), \mynote{and} $l_1$-ESPIRiT ($r=0.01$) were also performed.

\begin{figure}[t!]%
\centering%
  \includegraphics[width = 0.19 \textwidth]{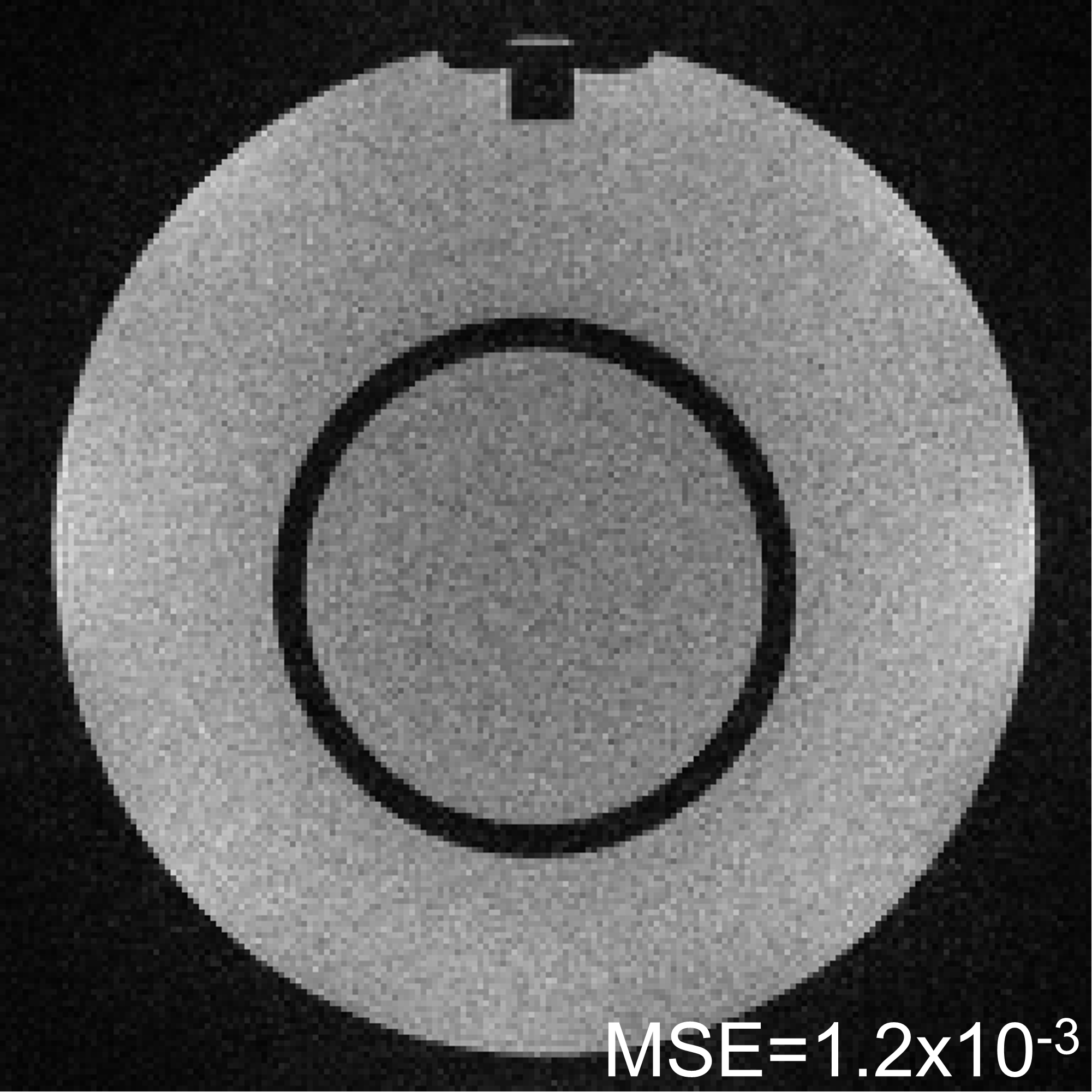}\hfill%
  \includegraphics[width = 0.19 \textwidth]{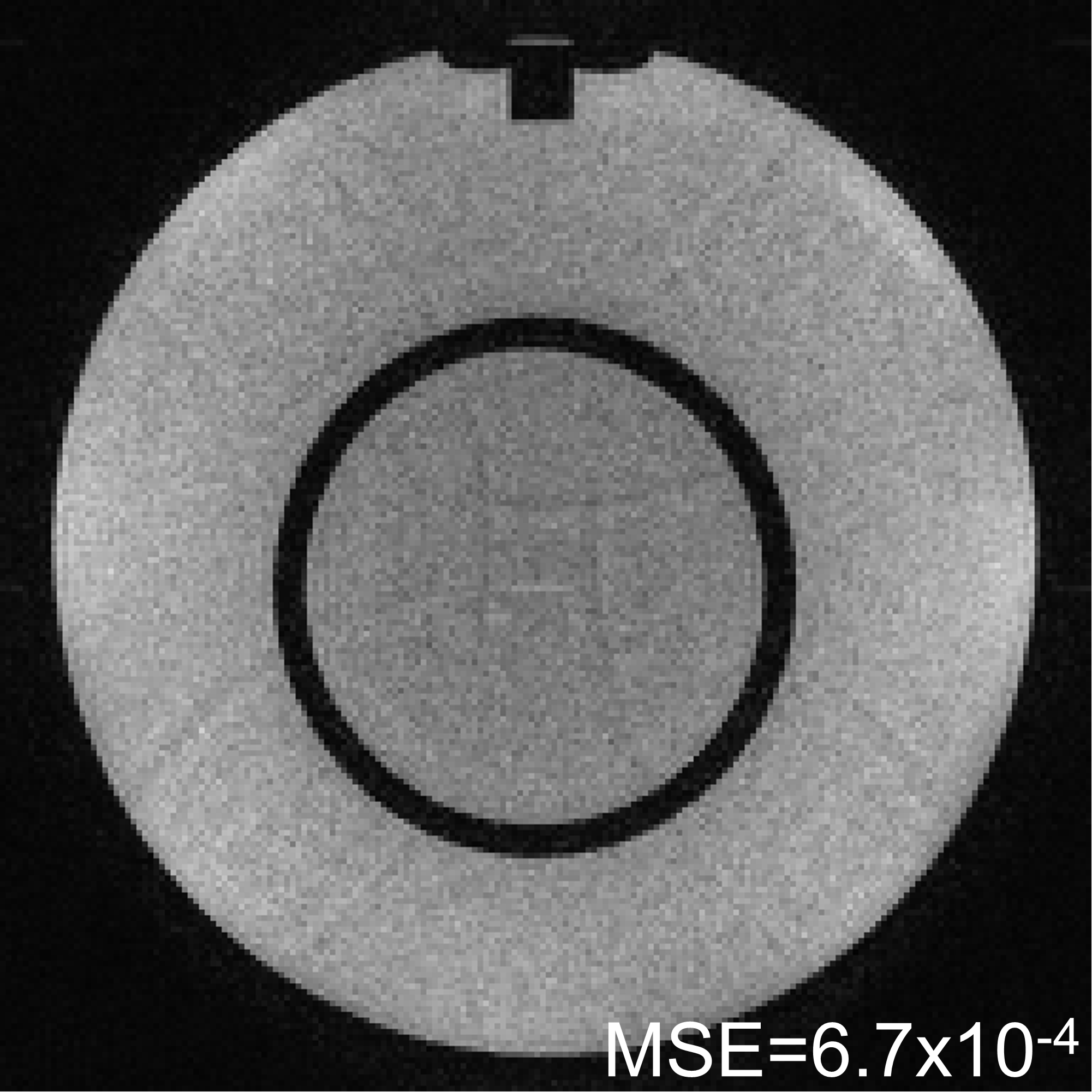}\hfill%
  \includegraphics[width = 0.19 \textwidth]{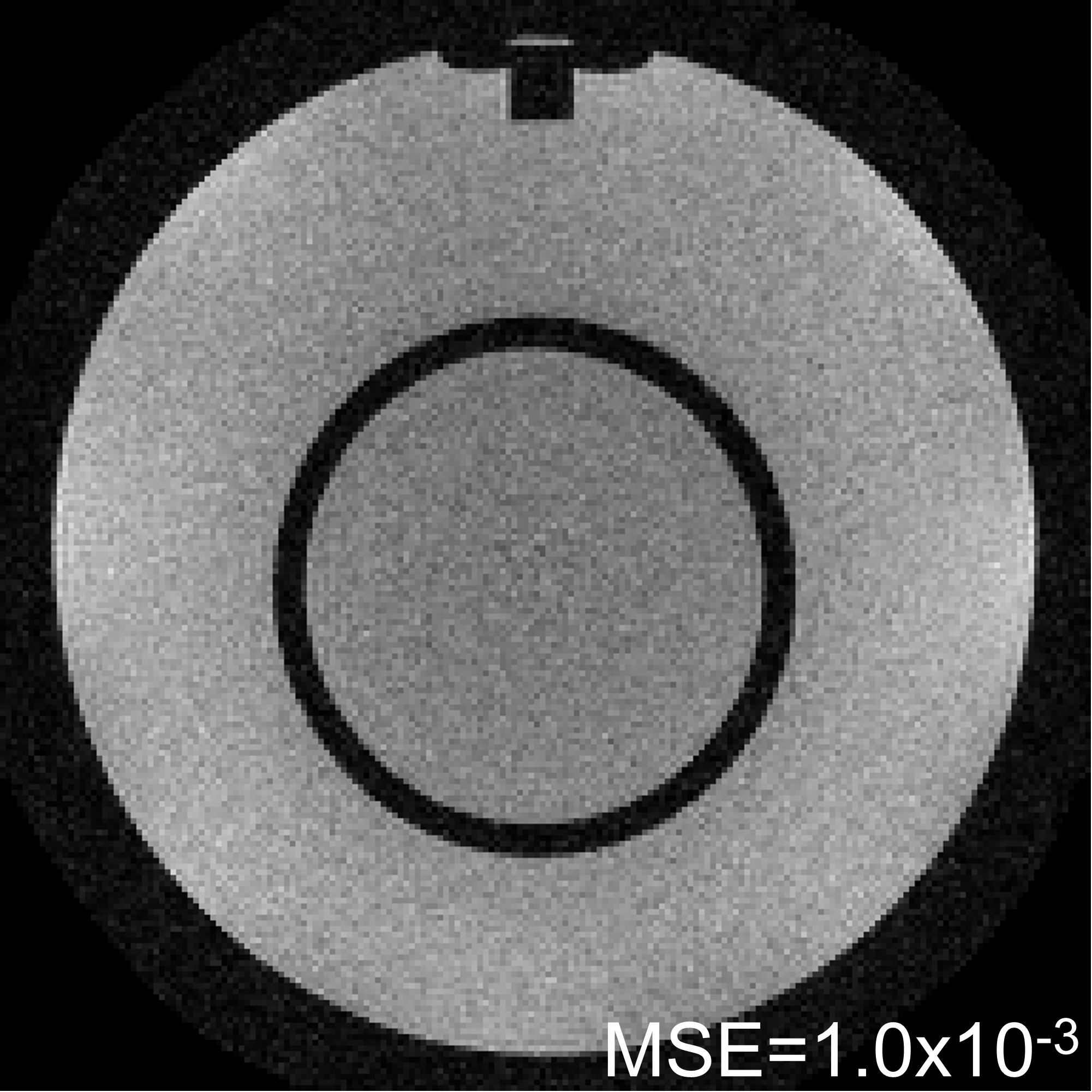}\hfill%
  \includegraphics[width = 0.19 \textwidth]{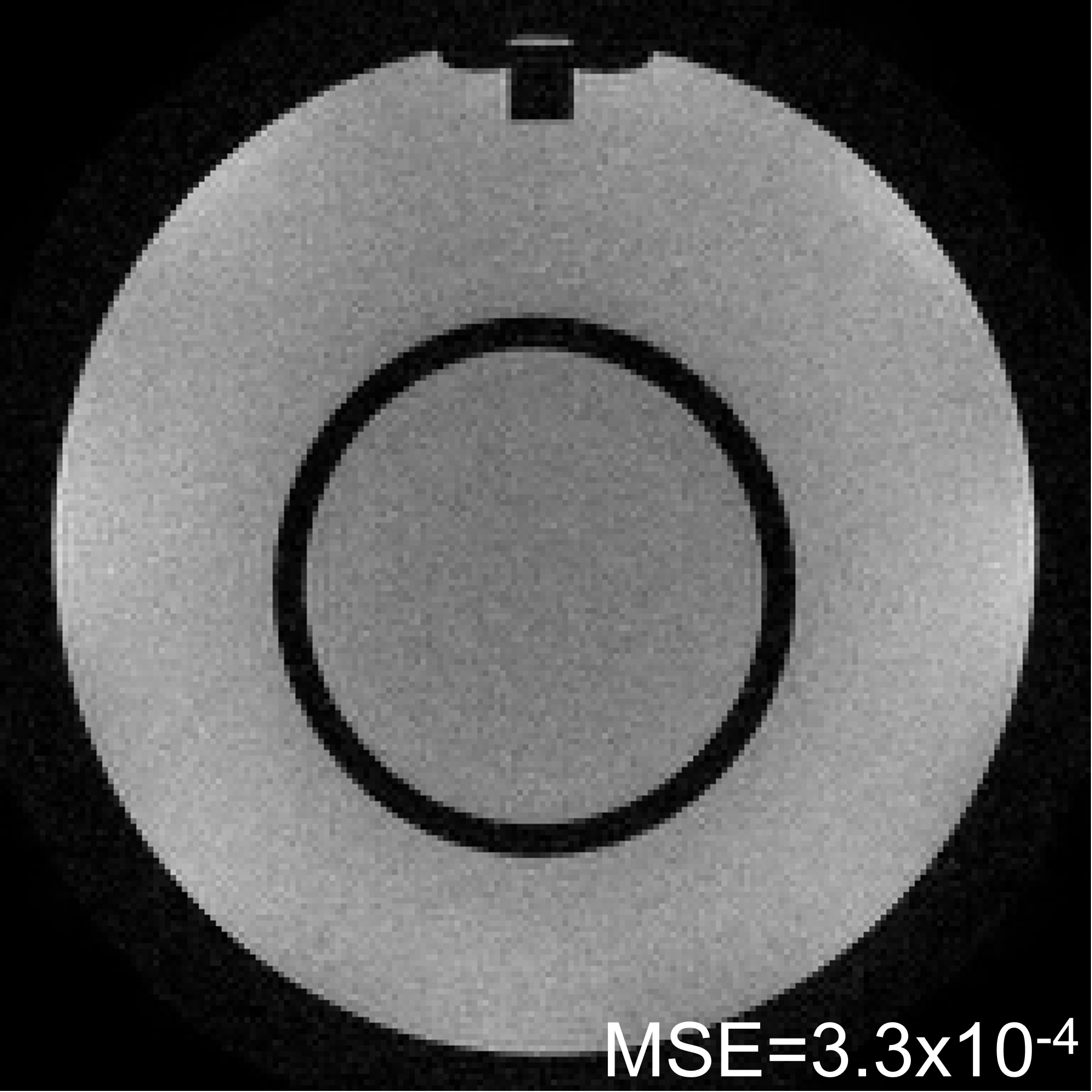}\hfill%
  \includegraphics[width = 0.19 \textwidth]{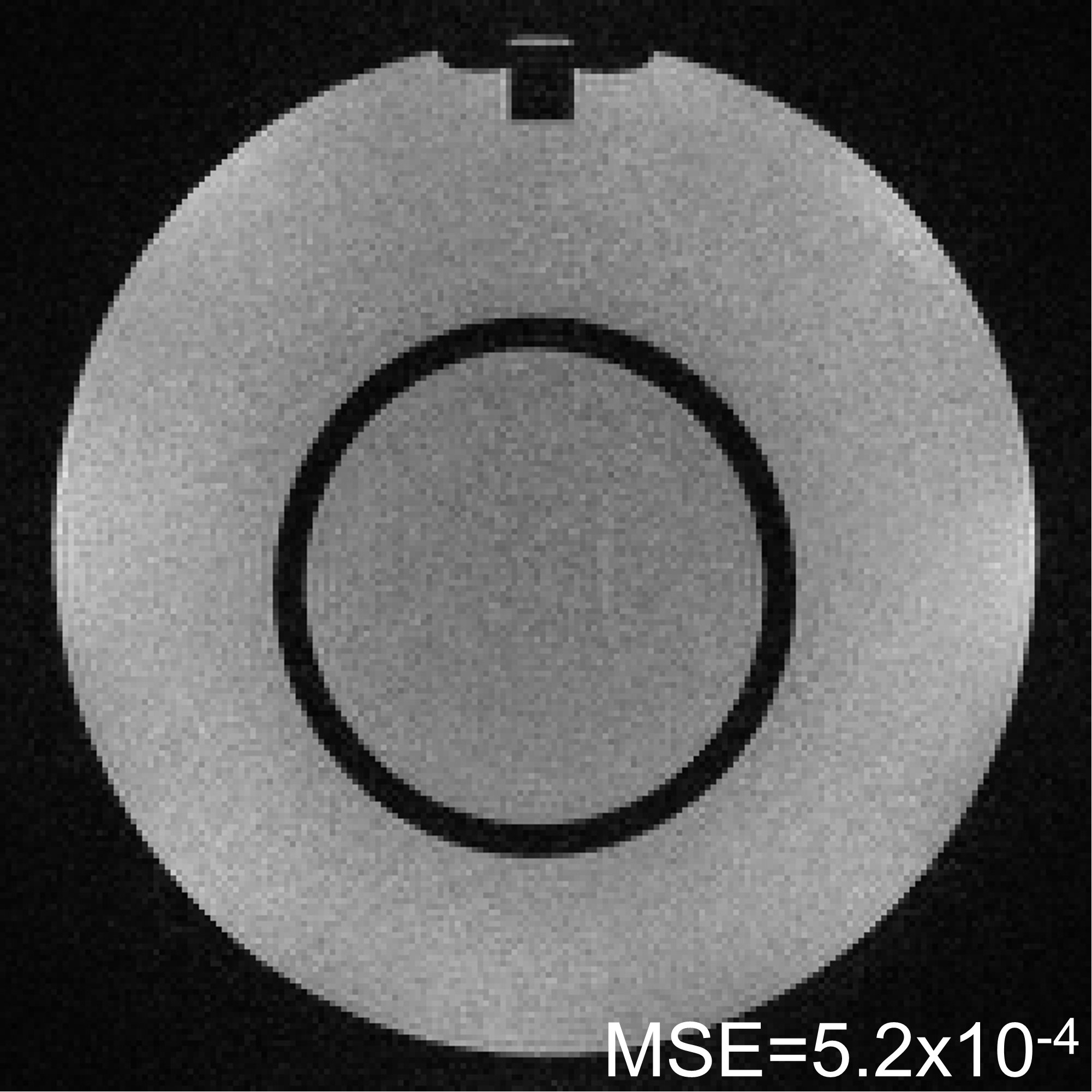}\hfill%
  \raisebox{4mm}{\includegraphics[width = 0.04 \textwidth]{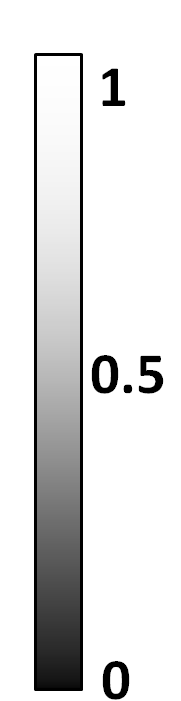}}\hfill%
\\[-0.05cm]%
  \includegraphics[width = 0.19 \textwidth]{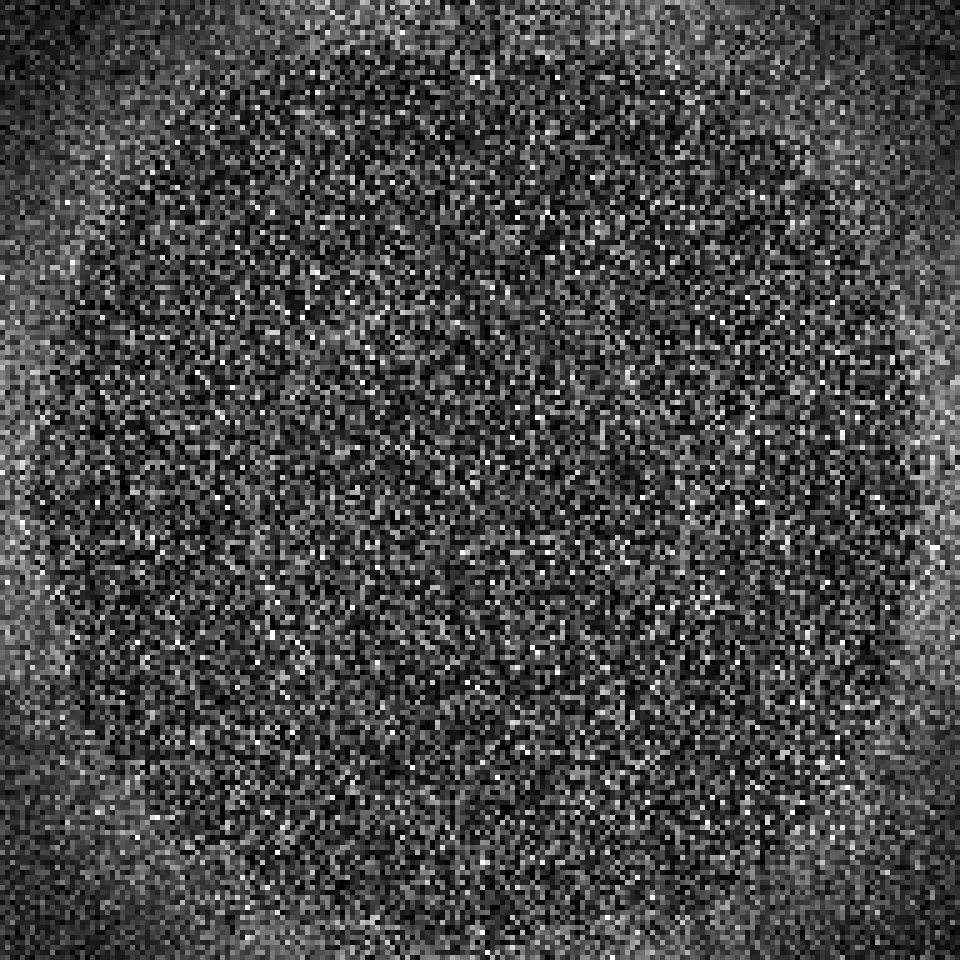}\hfill%
  \includegraphics[width = 0.19 \textwidth]{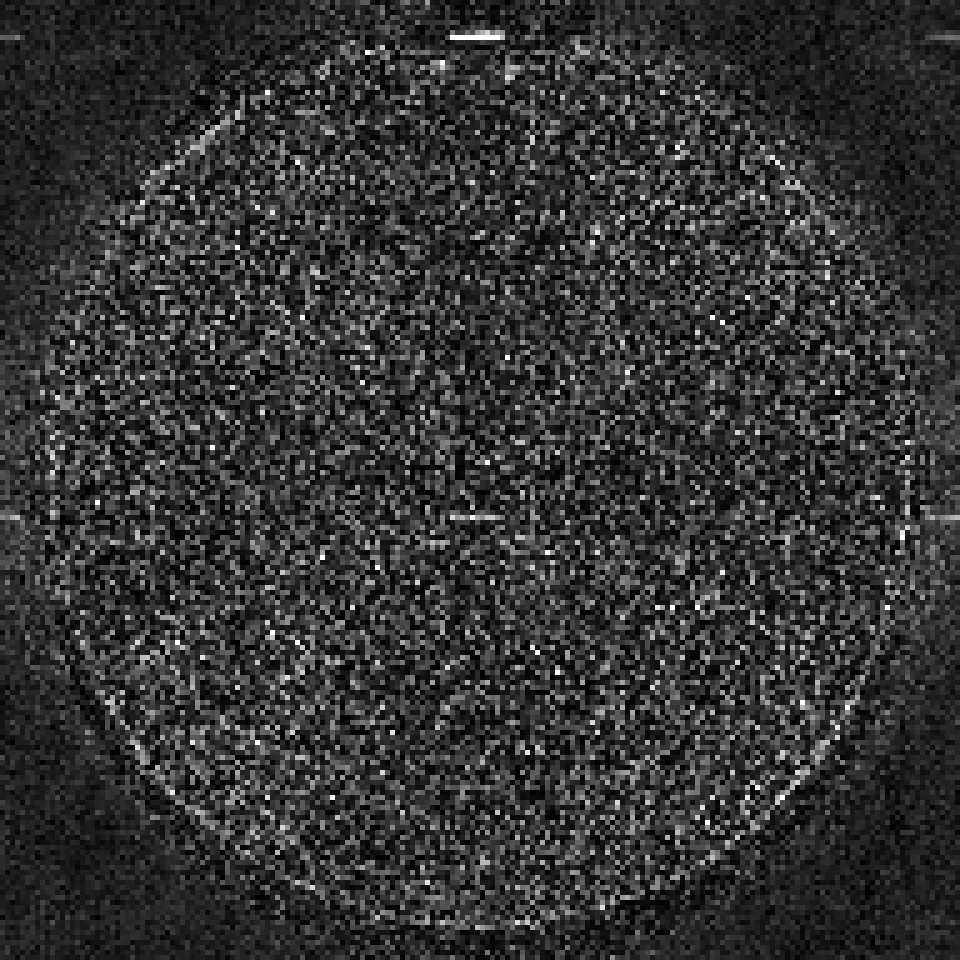}\hfill%
  \includegraphics[width = 0.19 \textwidth]{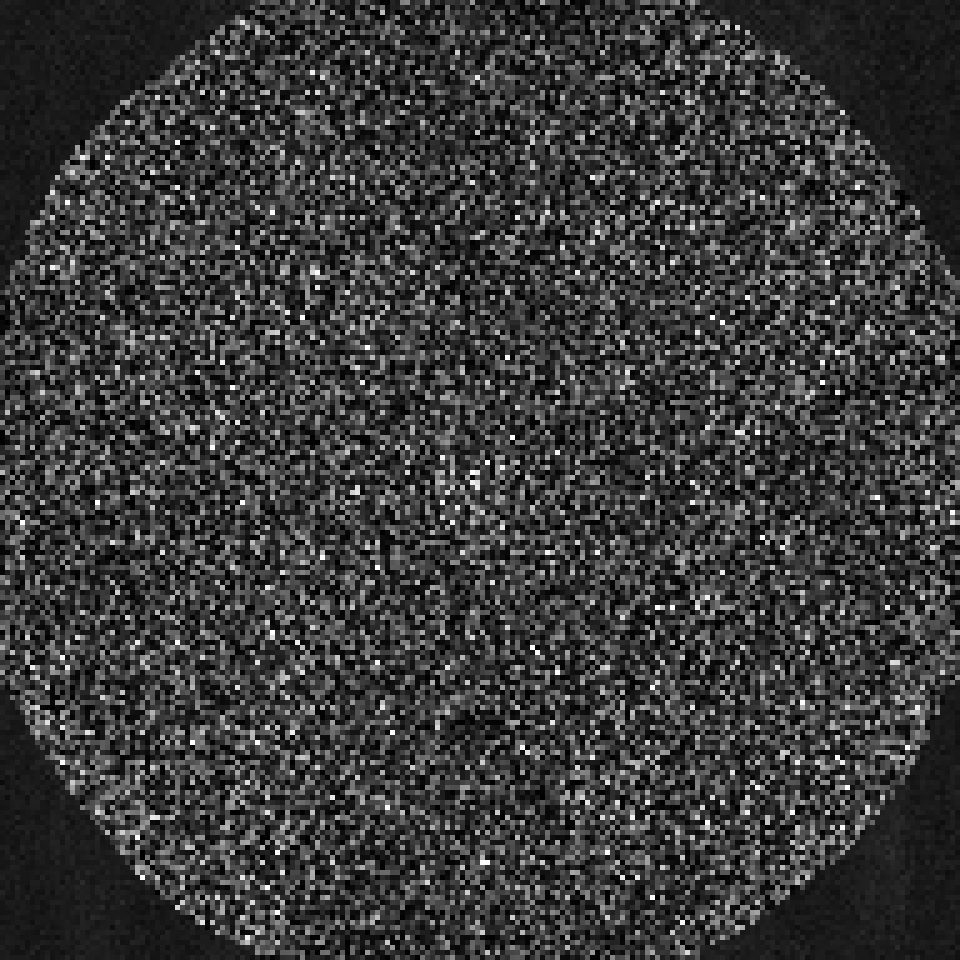}\hfill%
  \includegraphics[width = 0.19 \textwidth]{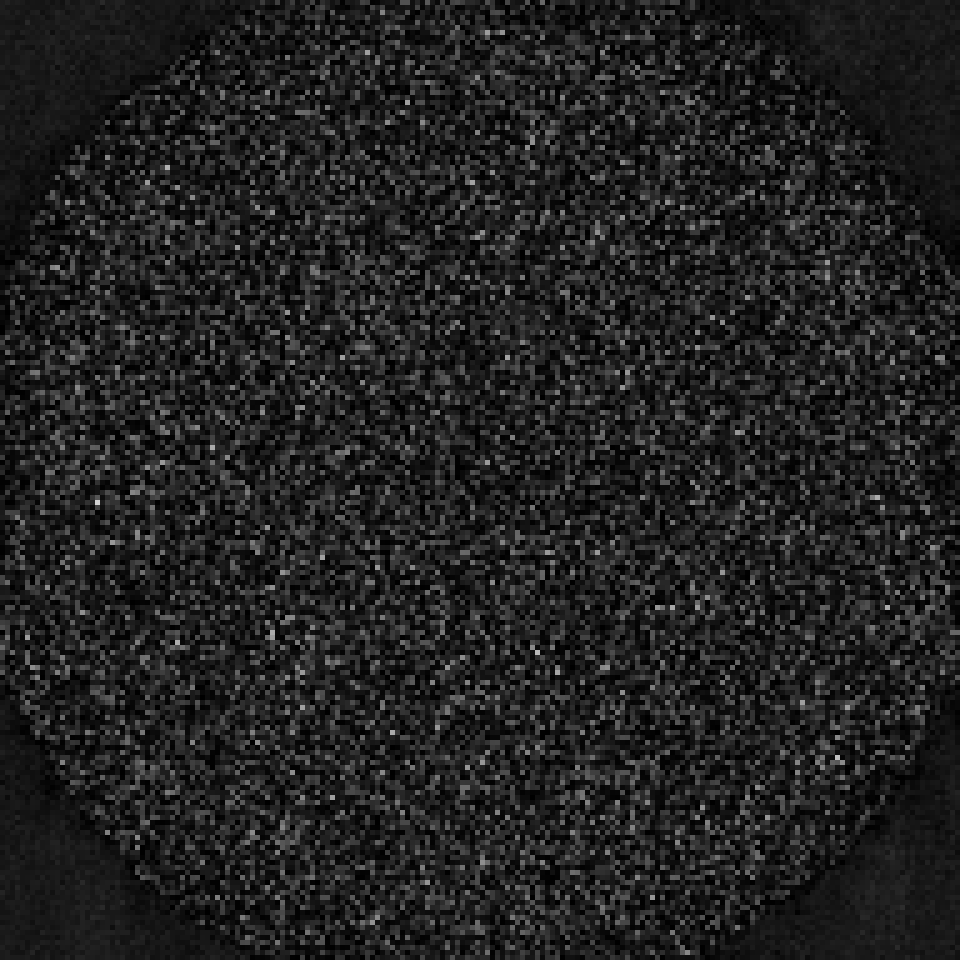}\hfill%
  \includegraphics[width = 0.19 \textwidth]{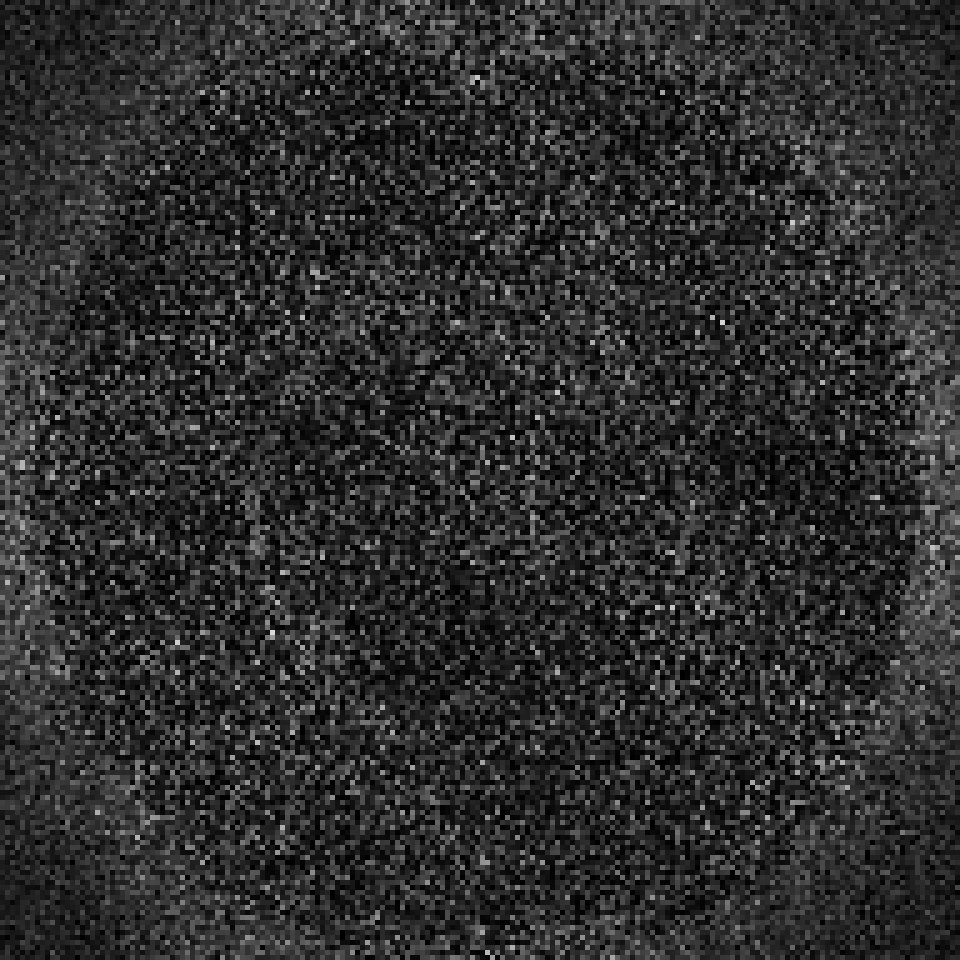}\hfill%
  \raisebox{4mm}{\includegraphics[width = 0.04 \textwidth]{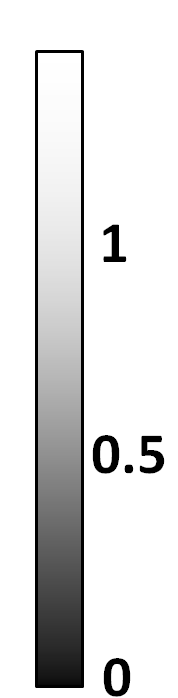}}\hfill%
\\[-0.19cm]%
  \subfigure[GRAPPA.]{\includegraphics[width = 0.19 \textwidth]{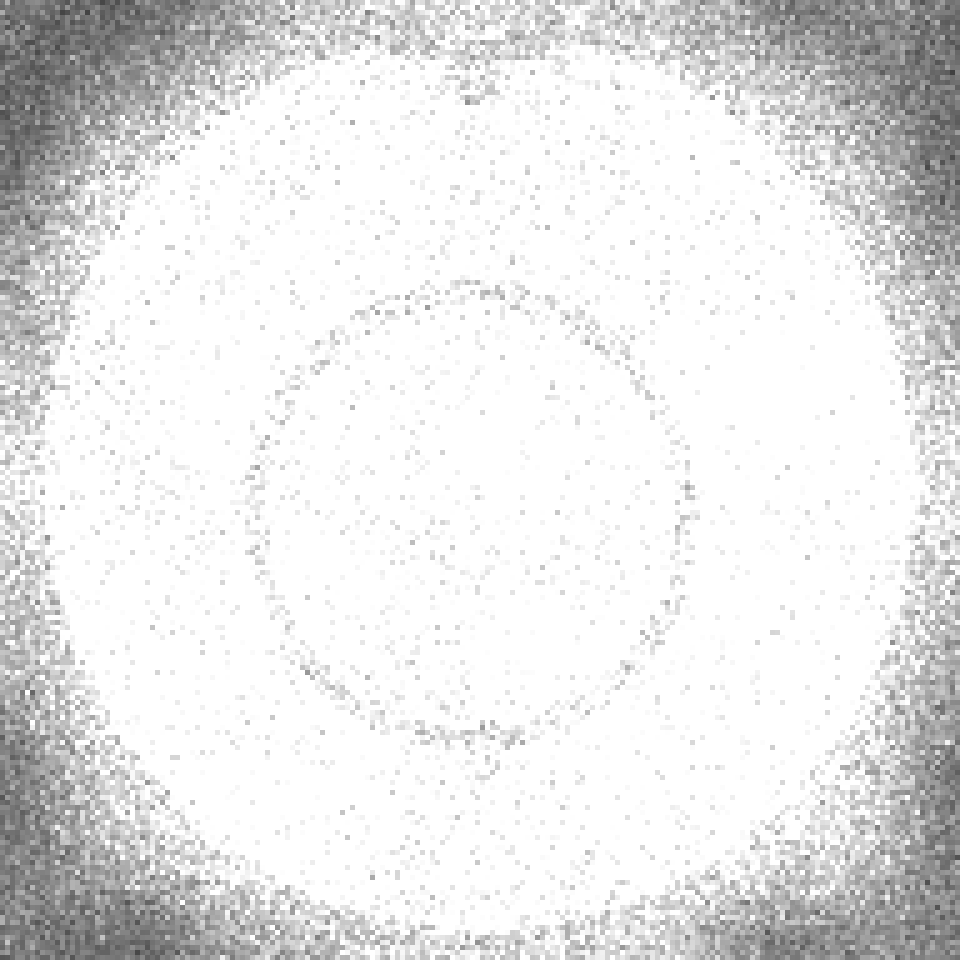}}\hfill%
  \subfigure[Regularized GRAPPA.]{\includegraphics[width = 0.19 \textwidth]{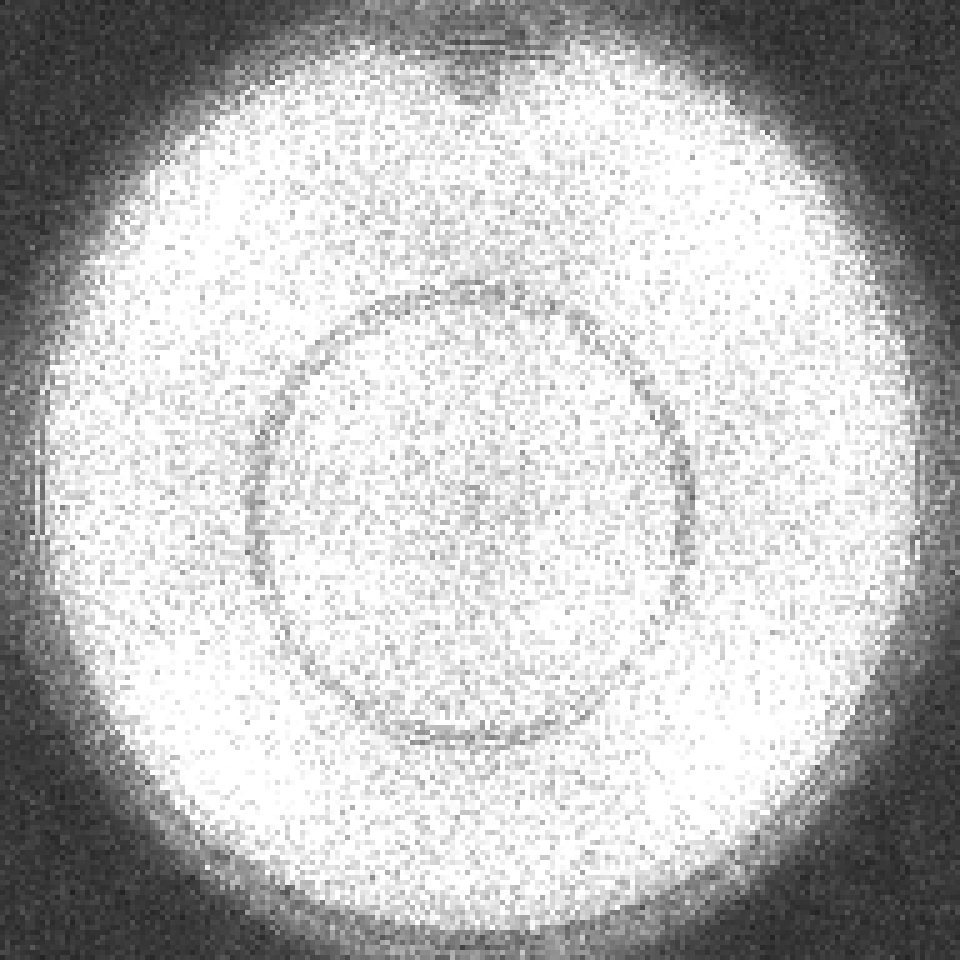}}\hfill%
  \subfigure[ESPIRiT.]{\includegraphics[width = 0.19 \textwidth]{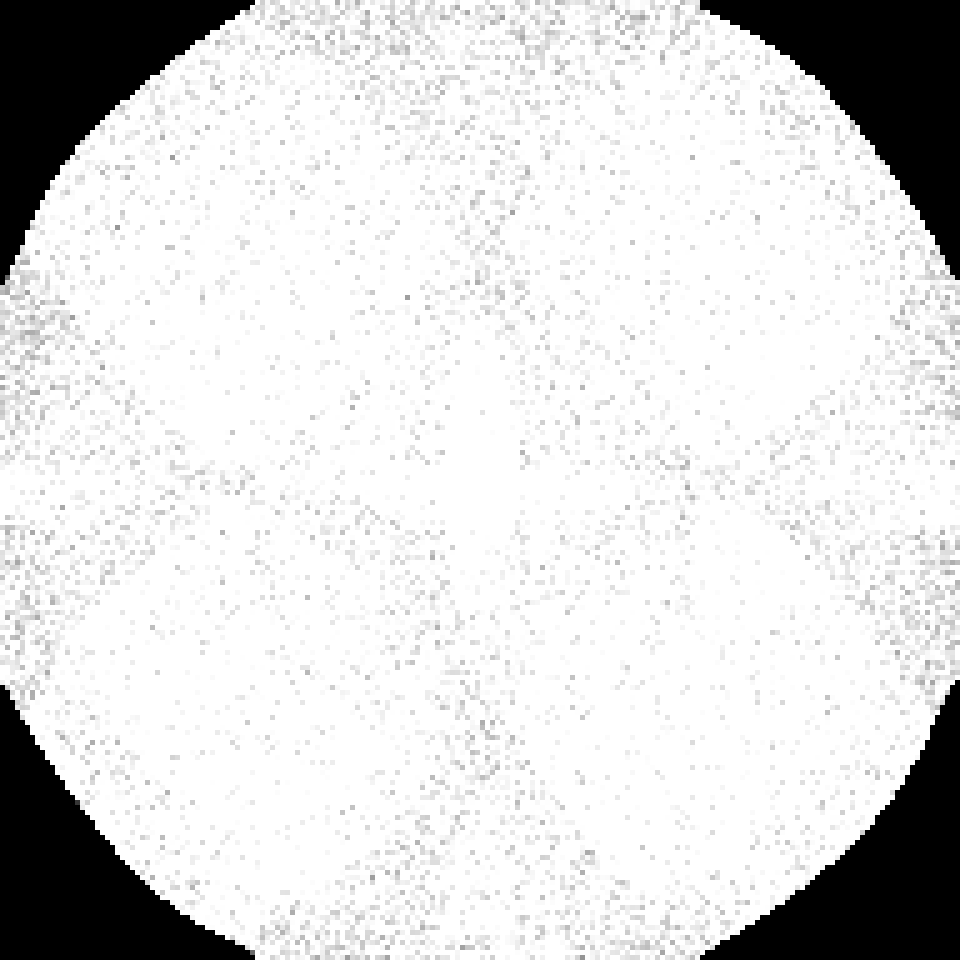}}\hfill%
  \subfigure[$l_1$-ESPIRiT.]{\includegraphics[width = 0.19 \textwidth]{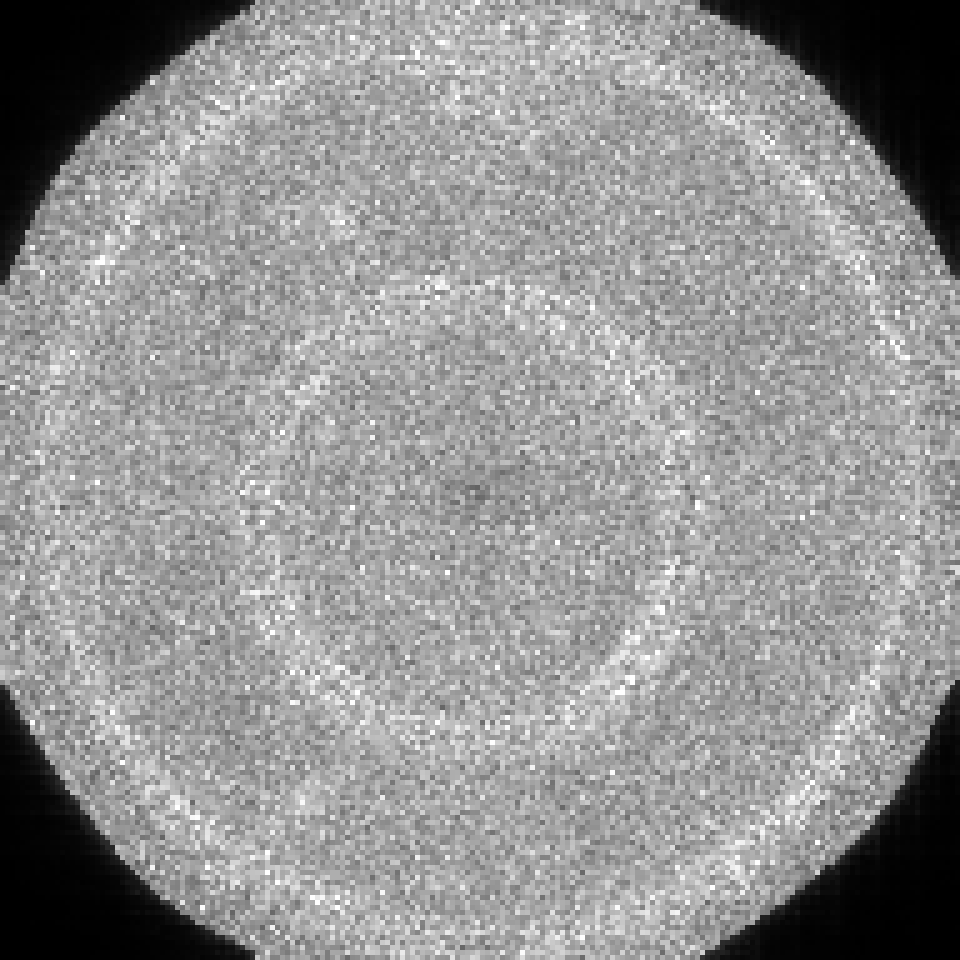}}\hfill%
  \subfigure[APIR-Net.]{\includegraphics[width = 0.19 \textwidth]{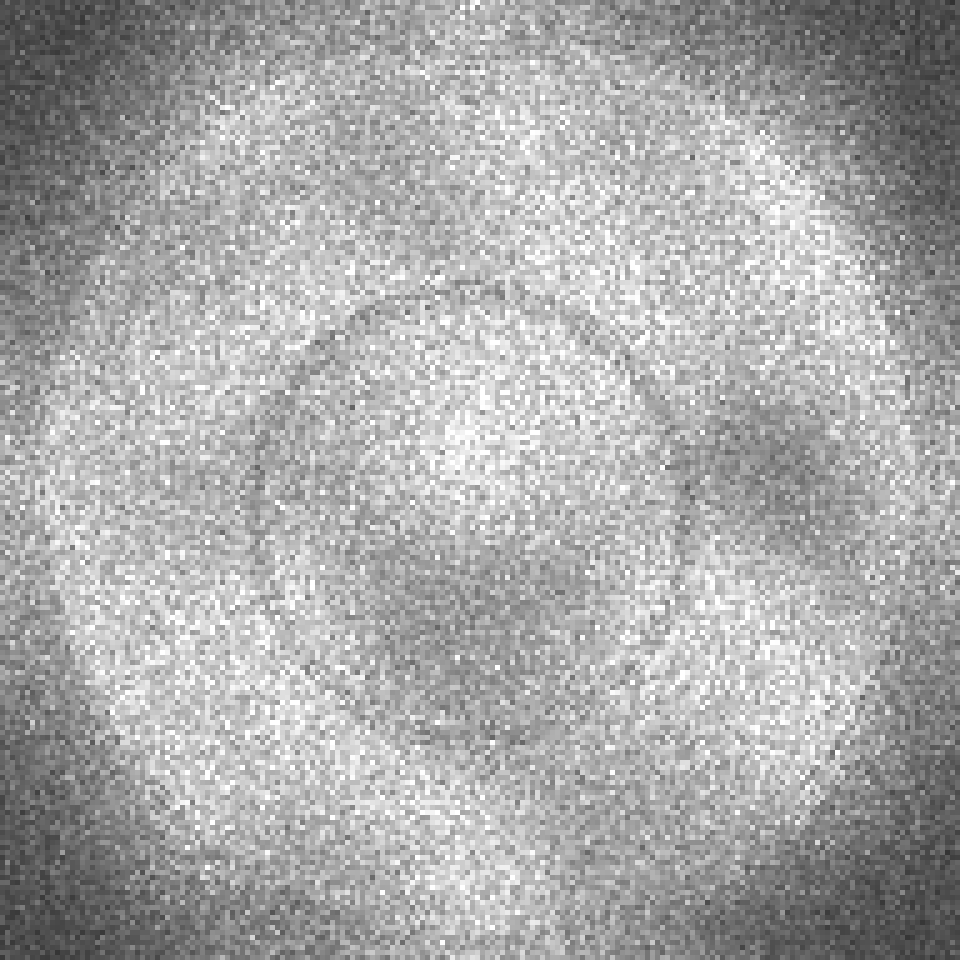}}\hfill%
  \raisebox{3mm}{\includegraphics[width = 0.04 \textwidth]{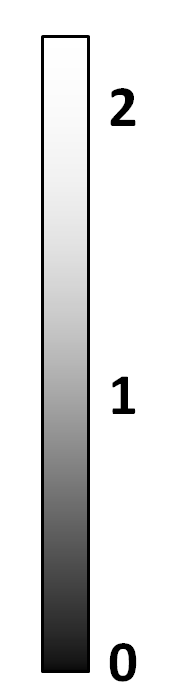}}\hfill%
\caption{One axial slice of the reconstructed images (first row), the reconstruction errors (second row), and the noise amplification factors (third row) of the phantom experiment.}
\label{Figure_phan}
\end{figure}
 
\section{Results}
\subsection{Evaluation with Phantom Acquisition}
The images of the phantom reconstructed by all methods are shown in the first row in Fig.~\ref{Figure_phan}. The reconstruction errors, i.e., the absolute difference between the reconstructed images and the reference image, are shown in the second row in Fig.~\ref{Figure_phan}. When using regularization with GRAPPA, SNR increases but aliasing artifacts start to appear.
APIR-Net reconstruction shows higher SNR than ESPIRiT reconstruction and the GRAPPA reconstruction without regularization, and less aliasing artifacts than the regularized GRAPPA reconstruction while having higher SNR. By integrating a properly weighted regularization term of $l_1$-norm of wavelet coefficients of the reconstructed image, $l_1$-ESPIRiT reduced noise level of ESPIRiT without raising visually obvious artifacts, and achieves the optimal image quality overall.
MSEs of the reconstructed images are shown in the reconstructed images in Fig.~\ref{Figure_phan}. APIR-Net reconstruction shows a lower MSE than the other methods except $l_1$-ESPIRiT. While $l_1$-ESPIRiT outperforms APIR-Net in the reconstruction quality, the same regularization, i.e. the sparsity constraint in wavelet transform, of $l_1$-ESPIRiT can be integrated in APIR-Net as well to improve its reconstruction quality.

As shown in the third row in Fig.~\ref{Figure_phan}, with regularization, the noise amplification was reduced in GRAPPA reconstruction, but still clearly higher than APIR-Net reconstruction. $l_1$-ESPIRiT overall shows the optimal noise amplification suppression with a substantial improvement over ESPIRiT.

\begin{figure}[t!]
\centering
  \subfigure[GRAPPA.]{\includegraphics[width = 0.20 \textwidth]{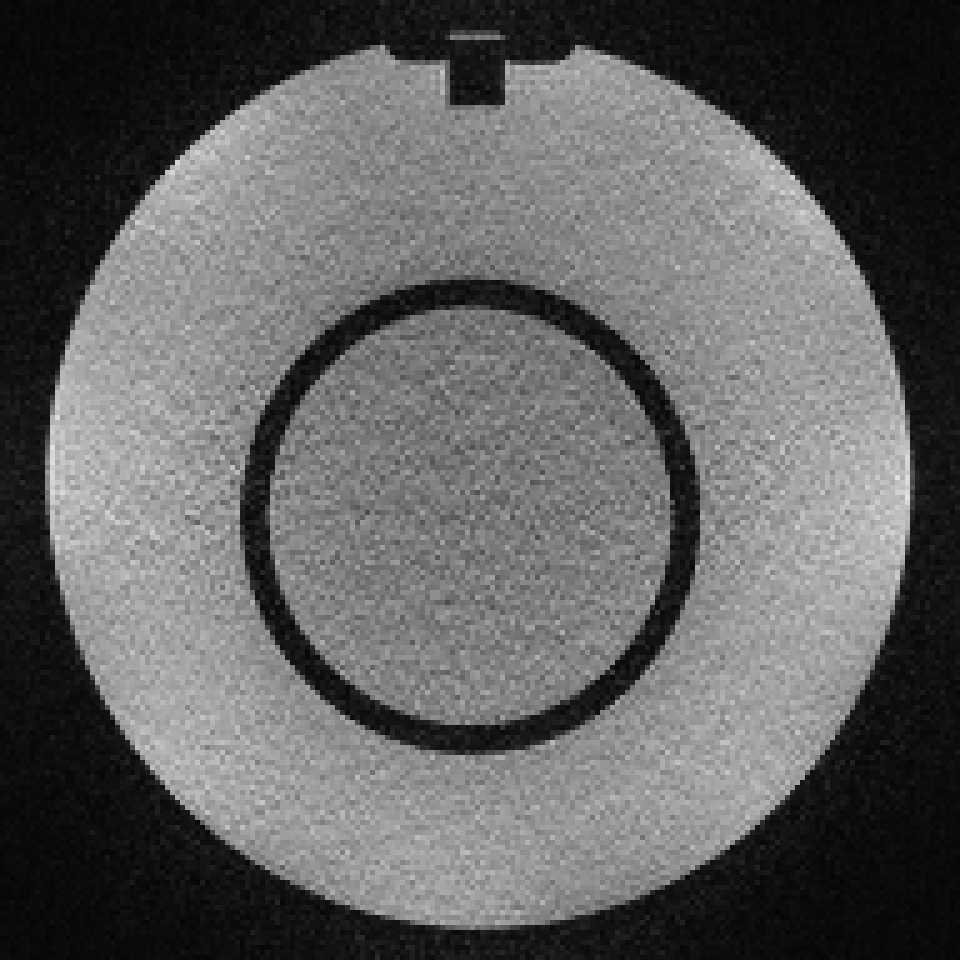}}
  \subfigure[Regularized GRAPPA.]{\includegraphics[width = 0.20 \textwidth]{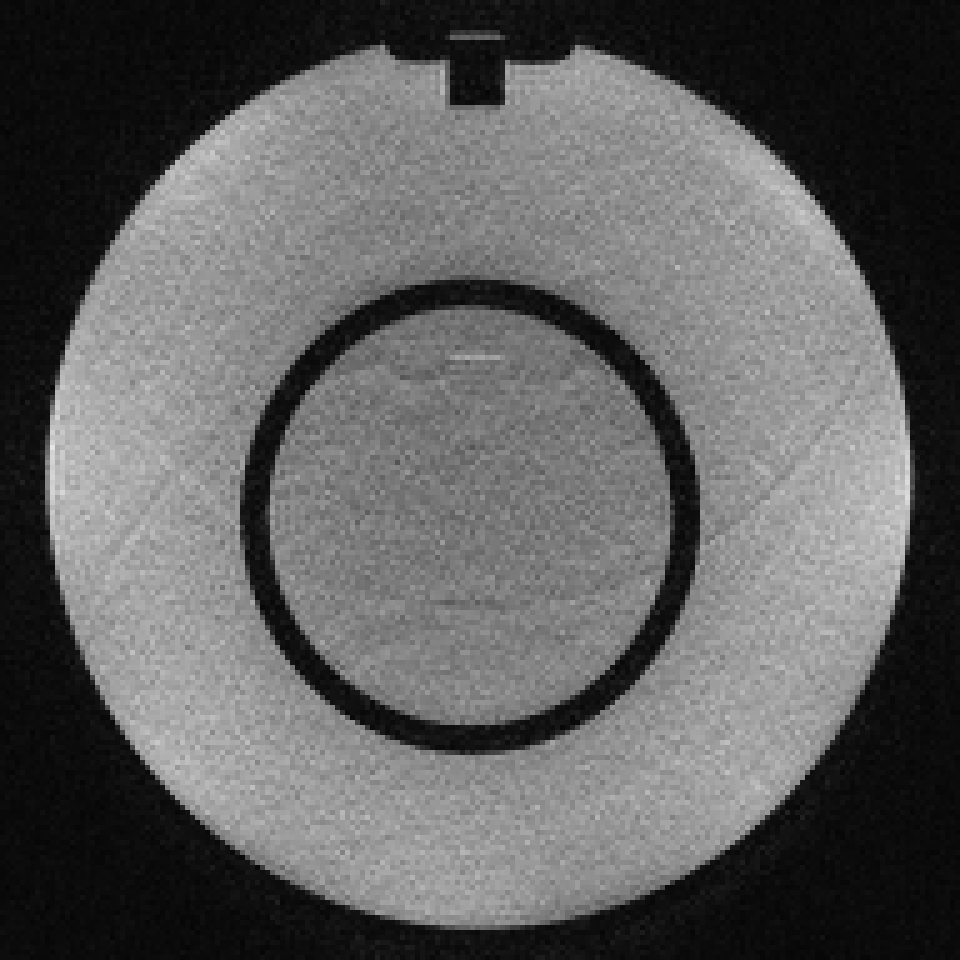}}
  \subfigure[ESPIRiT.]{\includegraphics[width = 0.20 \textwidth]{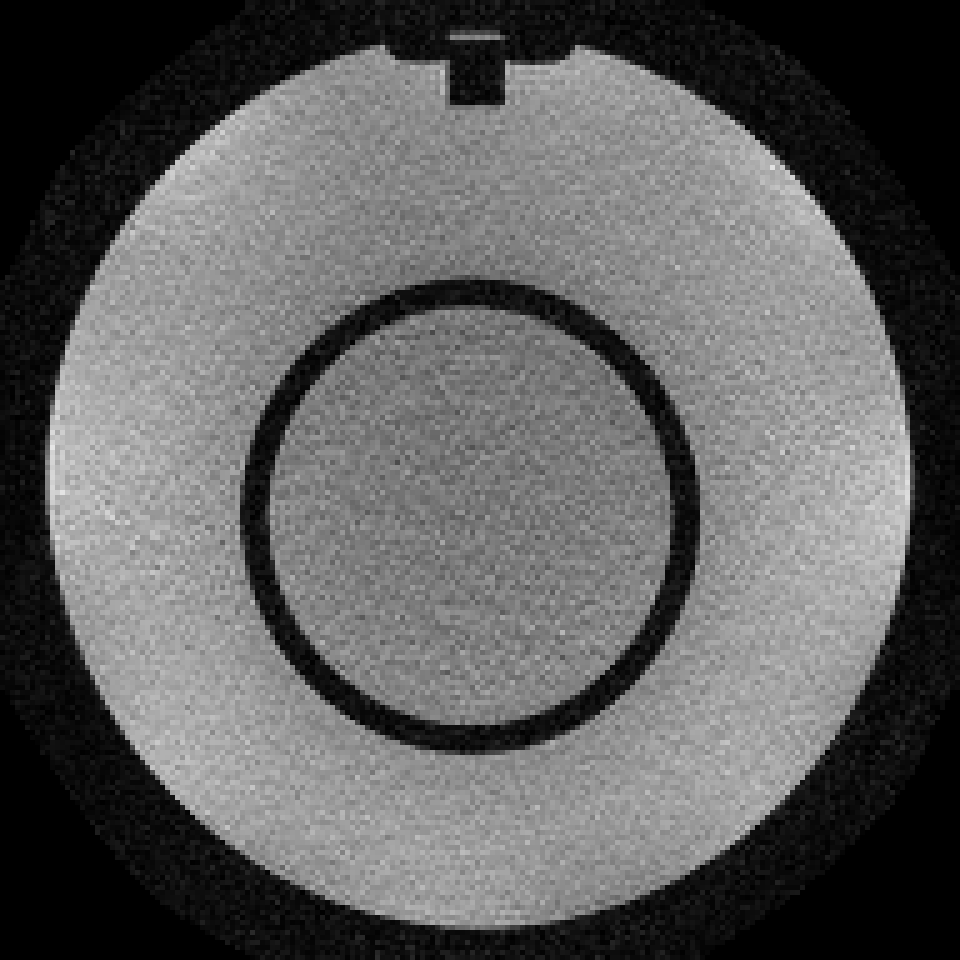}}
  \subfigure[$l_1$-ESPIRiT.]{\includegraphics[width = 0.20 \textwidth]{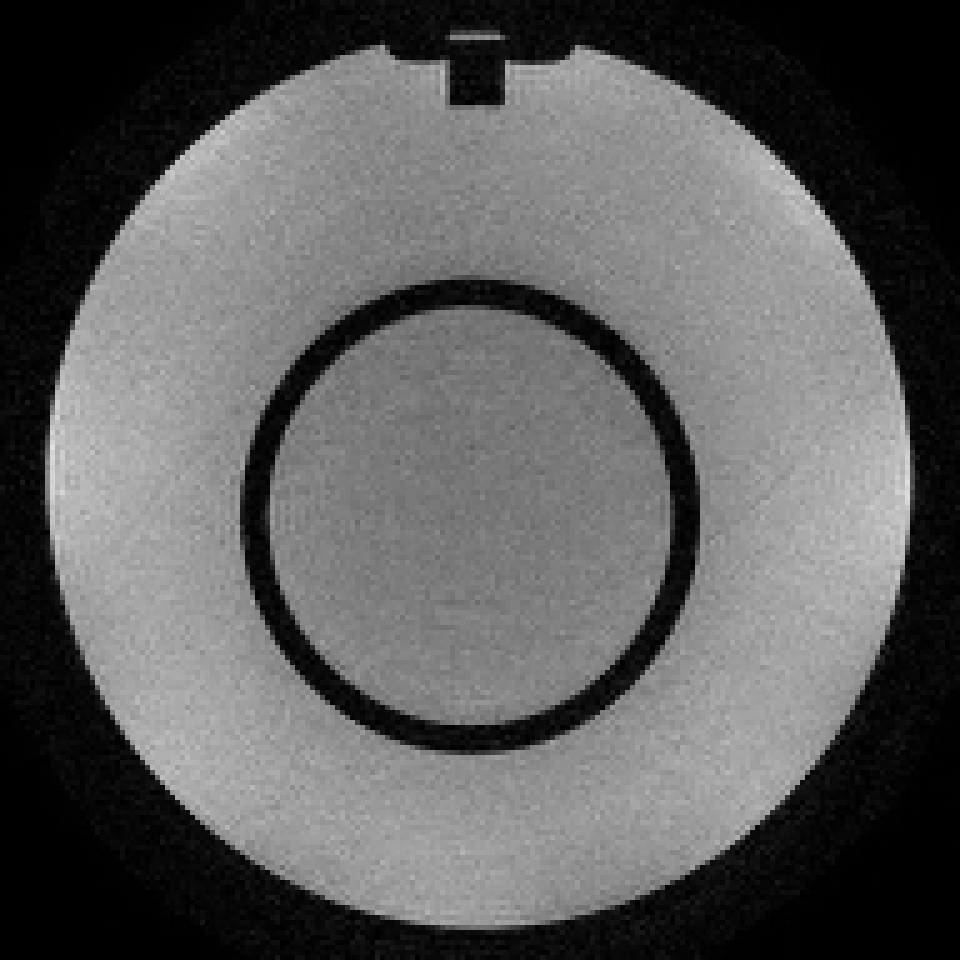}}
  \\
  \subfigure[RAKI.]{\includegraphics[width = 0.20 \textwidth]{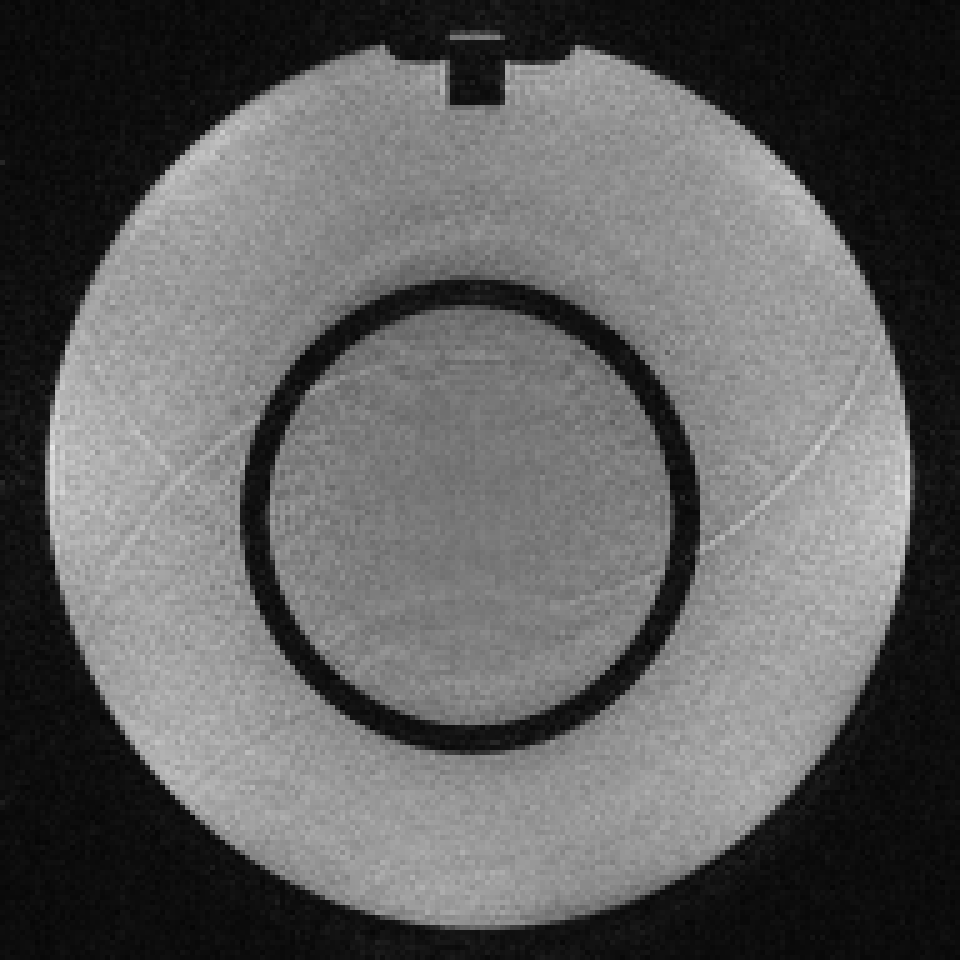}}
  \subfigure[2D APIR-Net.]{\includegraphics[width = 0.20 \textwidth]{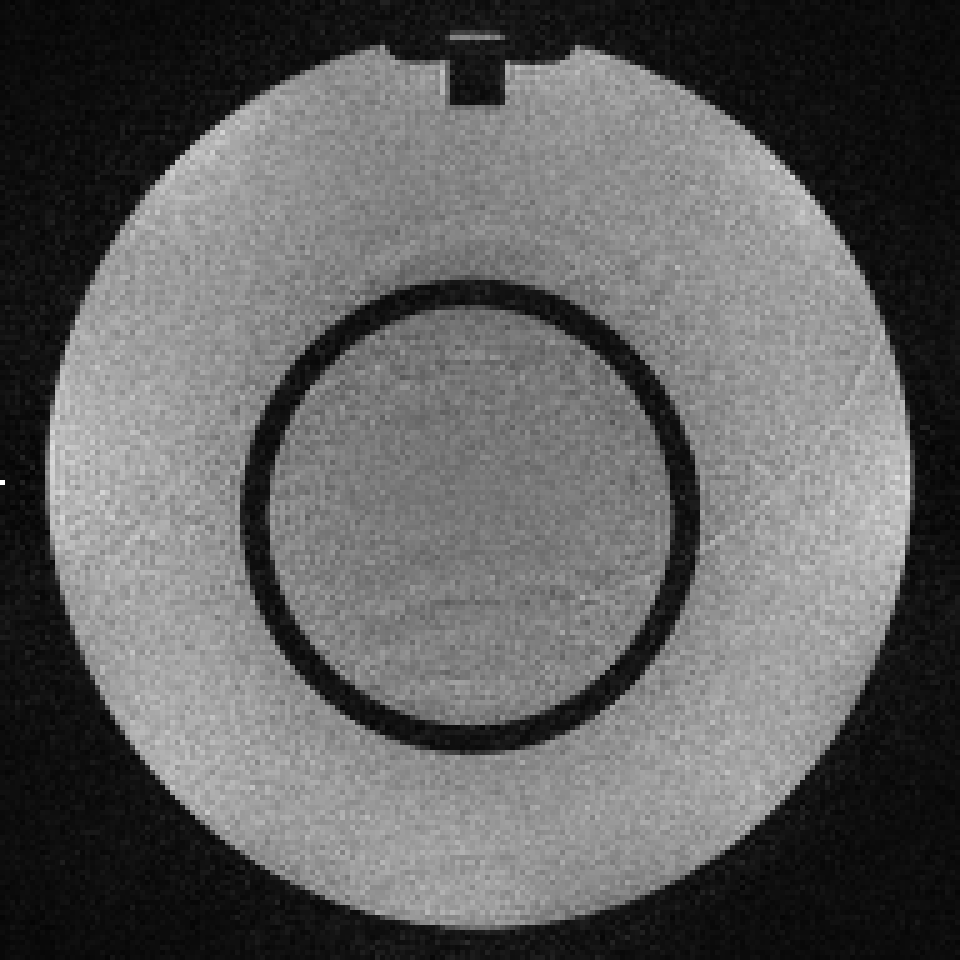}}
  \subfigure[3D APIR-Net.]{\includegraphics[width = 0.20 \textwidth]{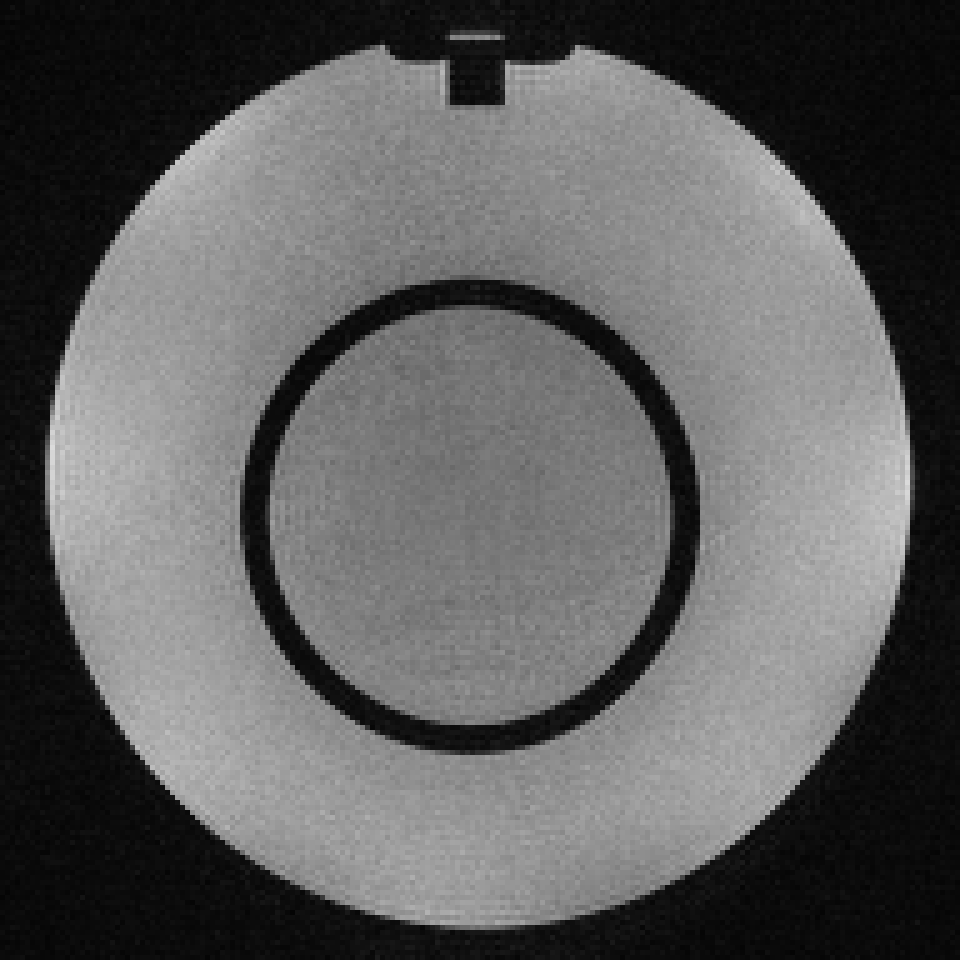}}
\caption{(a-f) 2D k-space reconstructions of one axial slice, (g) was reconstructed from a 3D k-space.}
\label{Figure_RAKI_comparison}
\end{figure}

\subsection{Comparison to RAKI}
As shown in Figure \ref{Figure_RAKI_comparison}, RAKI increased SNR substantially compared to GRAPPA. Compared to regularized GRAPPA, with similar amount of aliasing artifacts, RAKI still achieved higher SNR. With slightly lower SNR than RAKI and higher than regularized GRAPPA, 2D APIR-Net achieved less aliasing artifacts and visually better image quality. $l_1$-ESPIRiT substantially increased SNR compared to ESPIRiT, and achieved visually better quality than the previous ones. With more signals available for training, 3D APIR-Net shows improvement in both SNR and artifacts than 2D APIR-Net and also better image quality than other 2D methods.

\subsection{Evaluation with In-vivo Acquisitions}
\begin{figure}[t!]
\centering
  \subfigure[GRAPPA.]{\includegraphics[width = 0.20 \textwidth]{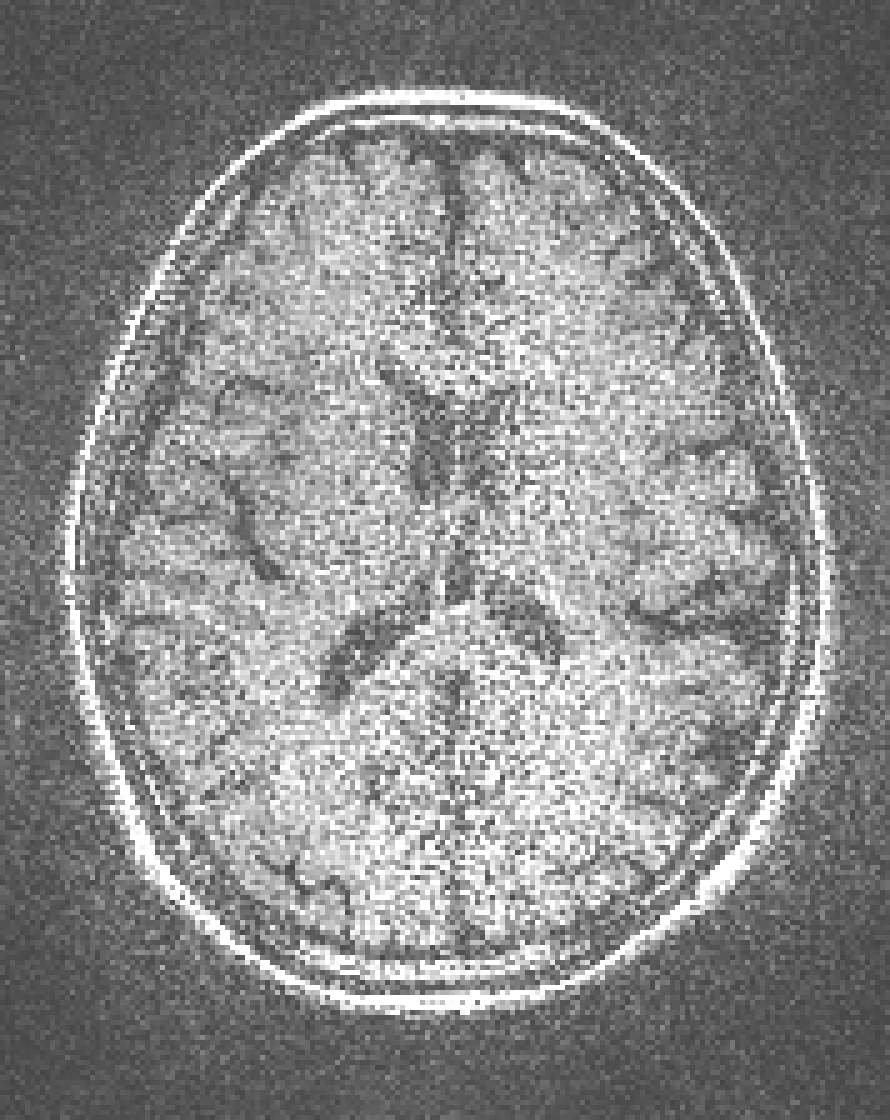}}\hfill%
  \subfigure[Regularized GRAPPA.]{\includegraphics[width = 0.20 \textwidth]{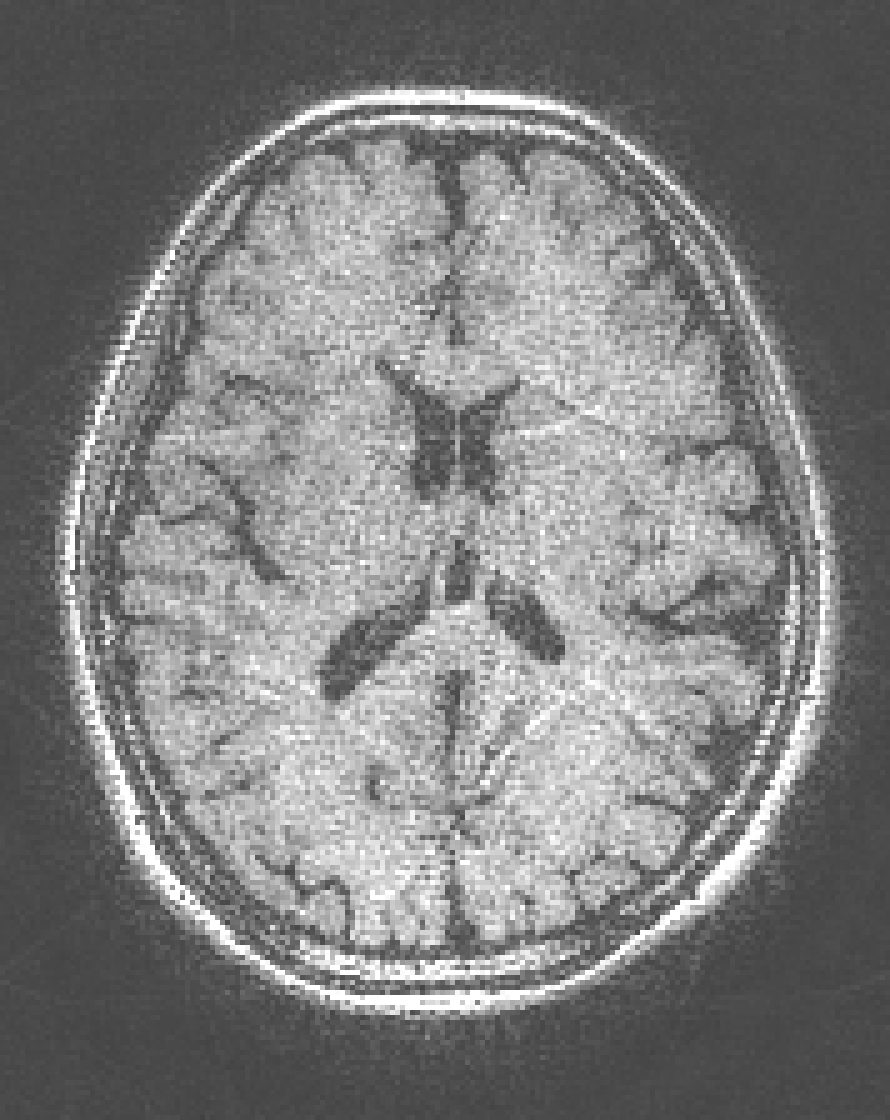}}\hfill%
  \subfigure[ESPIRiT.]{\includegraphics[width = 0.20 \textwidth]{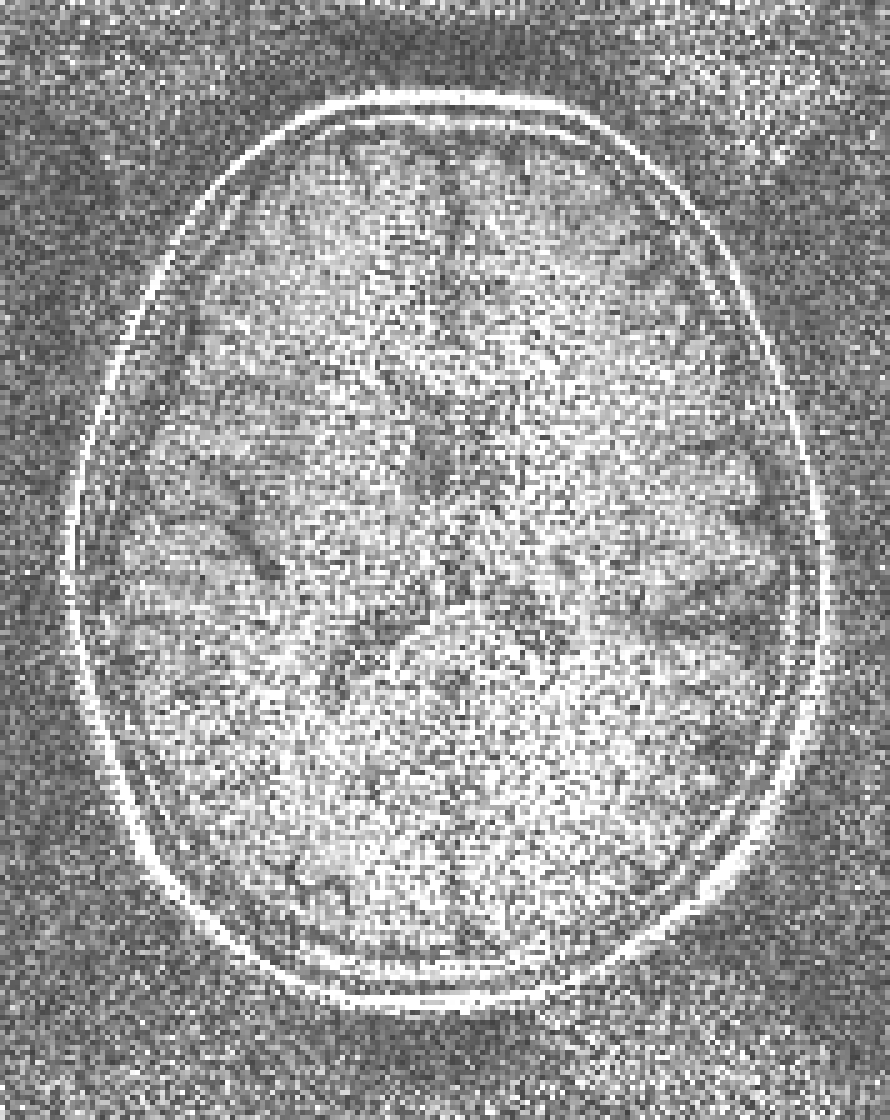}}\hfill
  \subfigure[$l_1$-ESPIRiT.]{\includegraphics[width = 0.20 \textwidth]{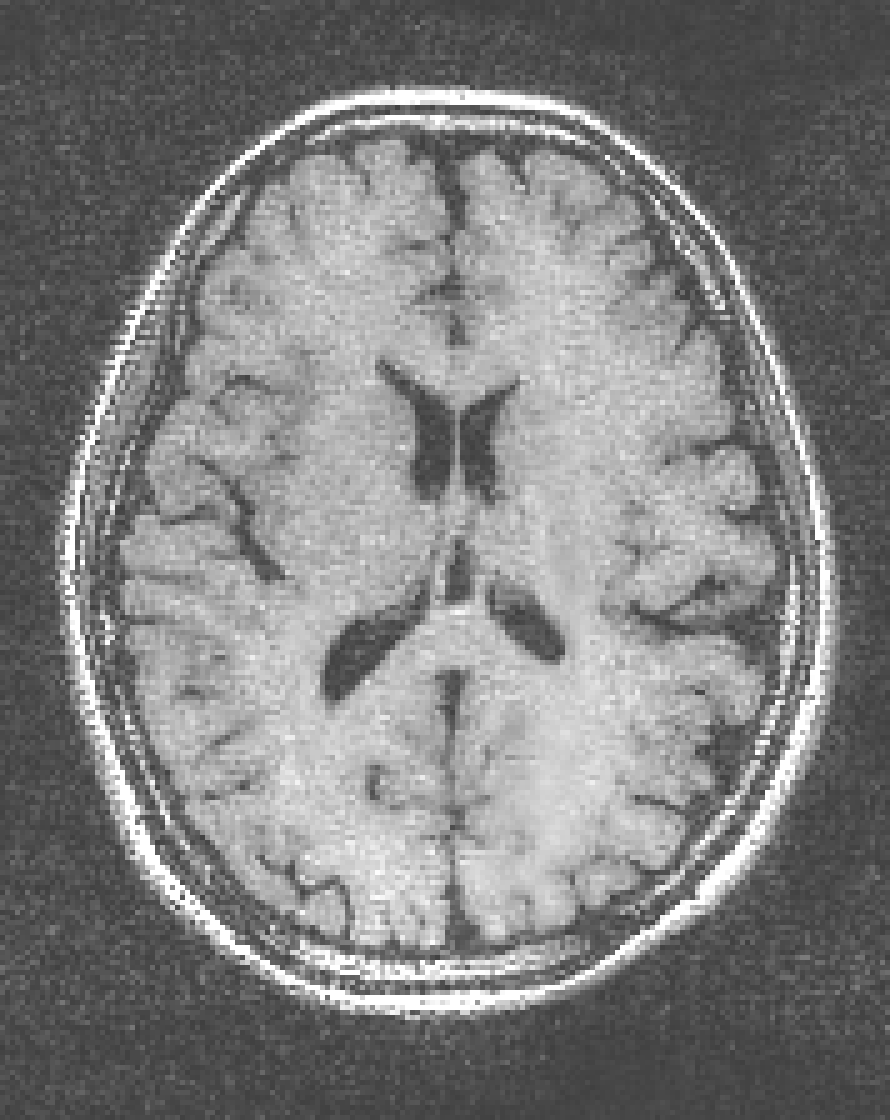}}\hfill
  \subfigure[APIR-Net.]{\includegraphics[width = 0.20 \textwidth]{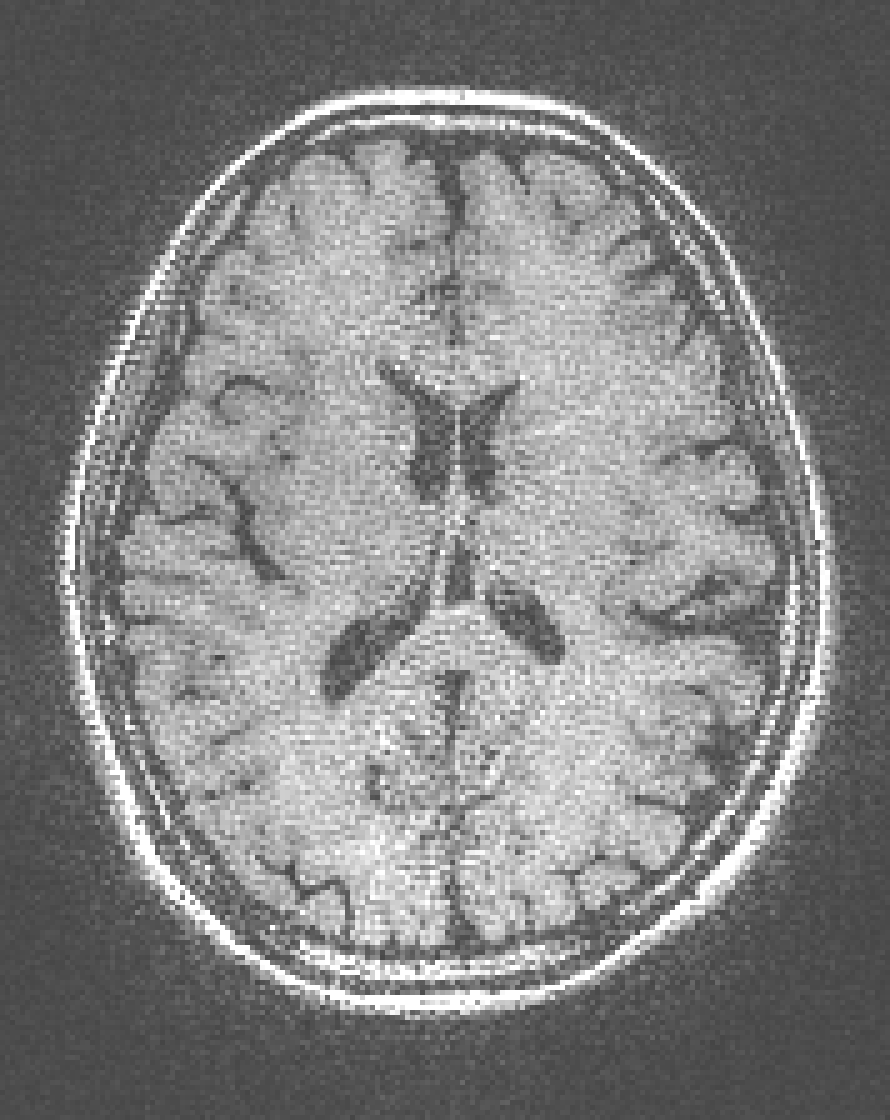}}\hfill
\caption{One axial slice of the reconstructed images of the prospectively subsampled in-vivo acquisition.}
\label{Figure_invivo}
\end{figure}

The reconstructed in-vivo images are shown in Fig.~\ref{Figure_invivo}. Regularized GRAPPA reconstruction shows reduced noise compared to GRAPPA without regularization, but aliasing artifacts appear. APIR-Net reconstruction achieves better performance than GRAPPA in both noise and aliasing artifacts. Compared to ESPIRiT, APIR-Net reduced the noise level of the image, though $l_1$-ESPIRiT achieves a further reduced noise level with slight blurring appears.

\begin{figure}[t!]
\centering
  \subfigure[Level 1.]{\includegraphics[width = 0.20 \textwidth]{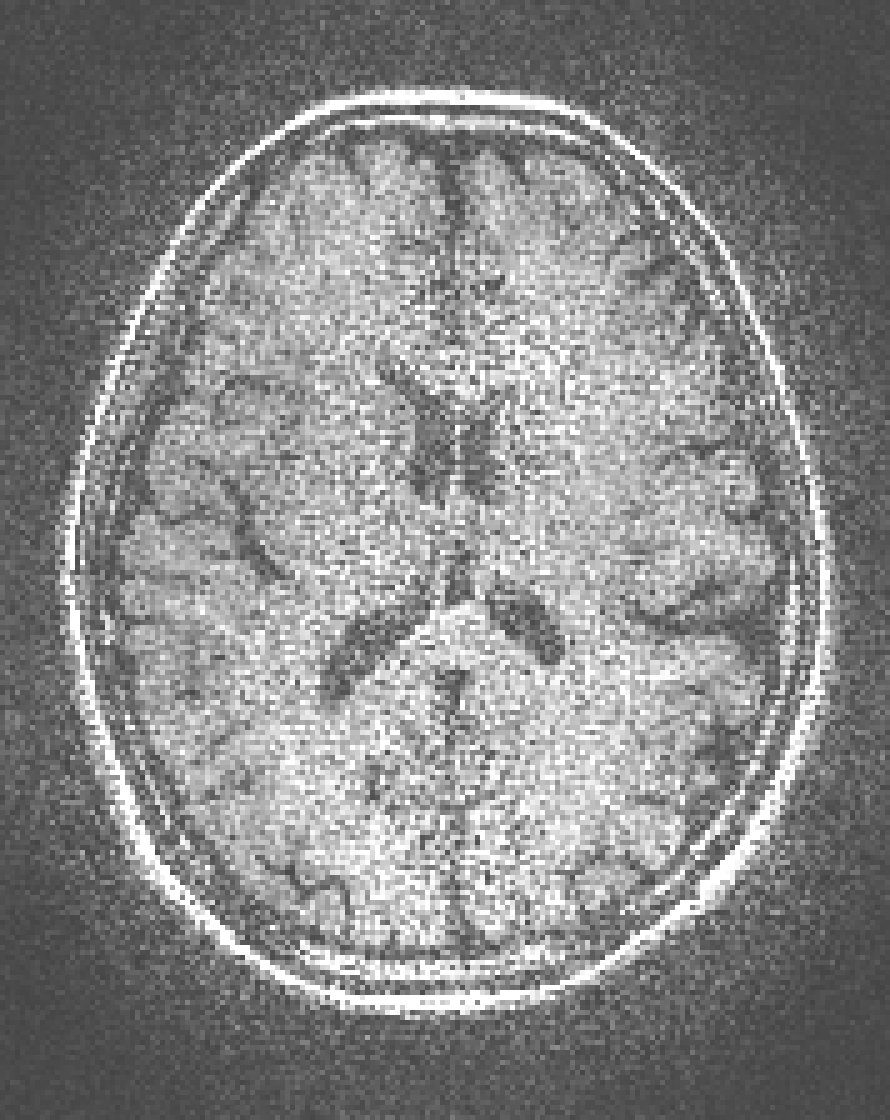}}
  \subfigure[Level 2.]{\includegraphics[width = 0.20 \textwidth]{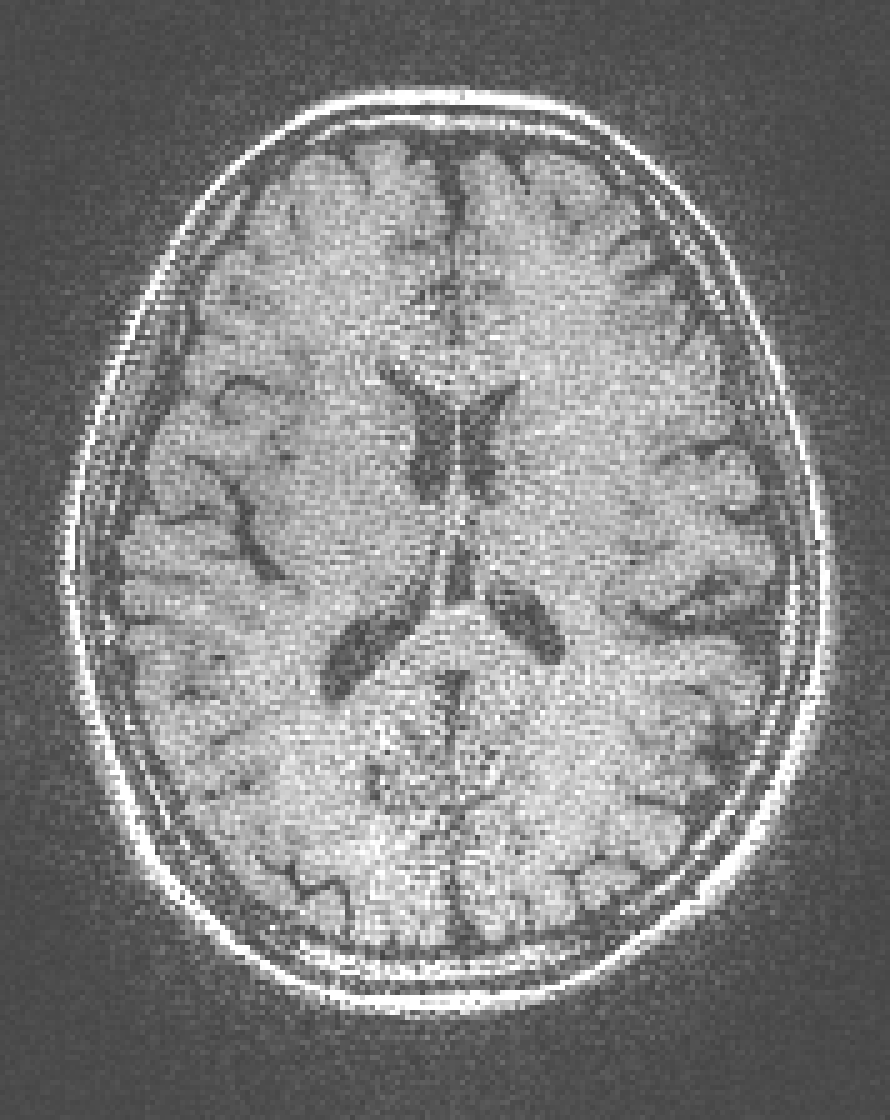}}
  \subfigure[Level 3.]{\includegraphics[width = 0.20 \textwidth]{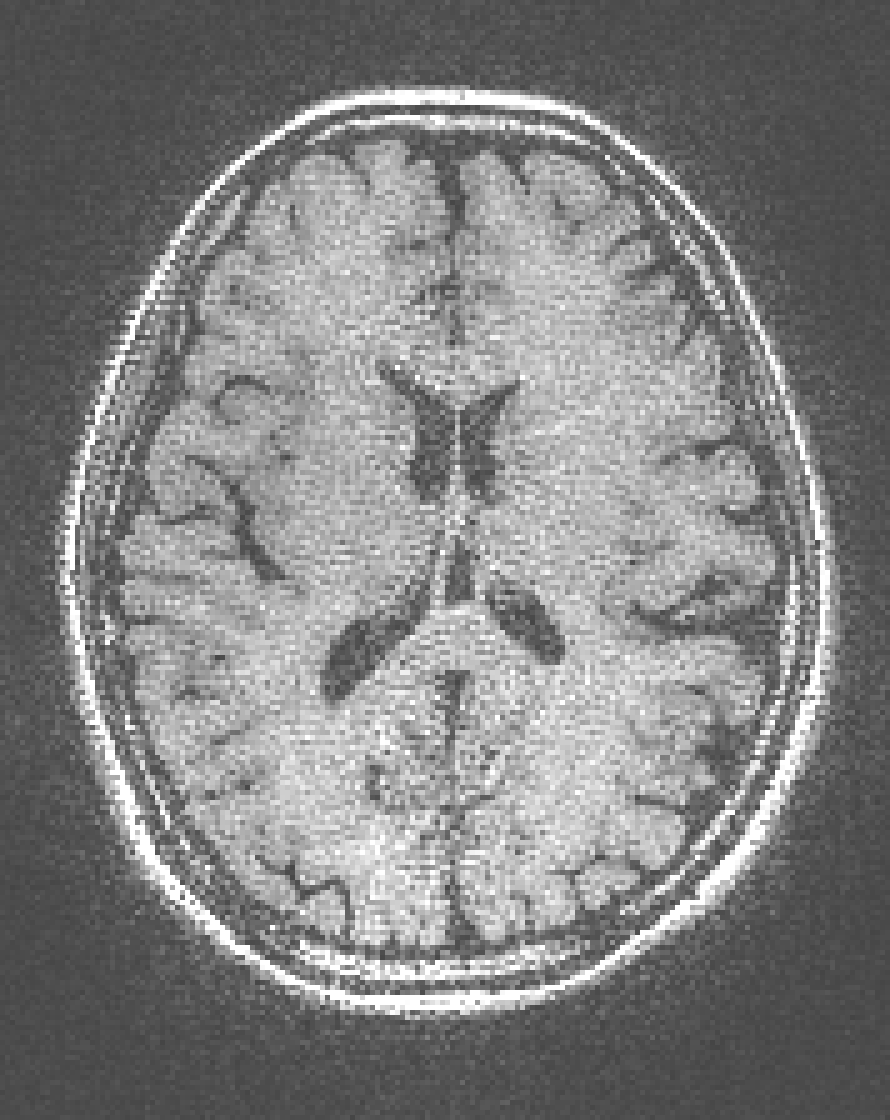}}
  \subfigure[Level 4.]{\includegraphics[width = 0.20 \textwidth]{UNET_FLAIR_full_23}}
\caption{One axial slice of the reconstructed in-vivo images by APIR-Net.
Reconstruction using weights of the first level training (a), second level (b), third level (c), and fourth level (d).}
\label{Figure_invivo_multiscales}
\end{figure}

Fig.~\ref{Figure_invivo_multiscales} shows the images reconstructed using weights trained from different levels in the hierarchical training. The image quality (in terms of noise and aliasing artifacts) is improved with the fine tuning of higher levels training.

\section{Discussion and Conclusion}
This work presented a novel method, APIR-Net, to reconstruct images from parallel imaging acquisitions using a neural network. While maintaining the non-linear optimization capability of deep learning based reconstruction methods, APIR-Net does not need representative training data of additional subject scans. This enables flexibility of APIR-Net in image size, contrast, or anatomy. 

Compared to GRAPPA, which estimates the unsampled signals in a linear way, APIR-Net achieves better noise amplification suppression, and thus better image quality and SNR. 
In GRAPPA noise amplification can be reduced by including Tikhonov regularization. However, this may introduce artifacts. In our results the regularized GRAPPA reconstruction had both stronger artifacts and higher noise level than APIR-Net.
Compared to ESPIRiT, \mynote{where excessive noise amplification exists}, APIR-Net shows improvement of the image quality with better SNR. 
$l_1$-ESPIRiT, which integrates the assumption of sparsity in the wavelet domain, substantially improves SNR and achieves (slightly) better image quality than APIR-Net, though tuning of the regularization strength is needed. As currently APIR-Net does not use any image based prior information, we hypothesise that the APIR-Net results can be further improved by including such prior information into the reconstruction, e.g. by adding the $l_1$ norm of the wavelet transform of the reconstructed image as additional cost term in equation \ref{APIR_Net_equation}.
The RAKI method \cite{akccakaya2019scan} which trains a convolutional neural network from ACS to predict the unsampled signals showed better results than GRAPPA. In APIR-Net, which uses a substantially different network architecture and extends RAKI in that all sampled signals (including signals beyond the ACS region) are used in prediction, improved image quality is achieved. 

Although it achieves the improved image quality, the current computation time of APIR-Net is much longer than GRAPPA. The high levels of the hierarchical training for APIR-Net are with typically very large size inputs (multi-channel high resolution 3D k-space), which makes it computationally expensive. We expect that by using patch generation techniques and stochastic optimization, the computation time can be reduced substantially. Additionally, initialization of the network weights might be improved by pretraining with previously acquired data; preferably with the same k-space pattern and receive coil. This may enable reducing the number of epochs and thus the computation time, while avoiding bias in reconstructed images due to the training dataset. 

To conclude, APIR-Net provides a promising alternative to the conventional parallel imaging methods, and results in improved image quality especially for low SNR acquisitions.
%
%
%
%
%
 \bibliographystyle{splncs04}
 \bibliography{jan_abbr,chaoping}
%
%
%
%
%
\end{document}